\documentclass{article}

\usepackage{arxiv}

\usepackage[utf8]{inputenc} 
\usepackage[T1]{fontenc}    
\usepackage[hidelinks]{hyperref}       
\usepackage{url}            
\usepackage{booktabs}       
\usepackage{amsfonts, amsmath}       
\usepackage{amsthm}
\usepackage{nicefrac}       
\usepackage{microtype}      
\usepackage{natbib}
\usepackage{doi}
\usepackage{fancyhdr, graphicx, float, geometry}
\usepackage{enumitem}
\usepackage{tikz}
\usepackage{subcaption}

\newtheorem{remark}{Remark}
\newtheorem{theorem}{Theorem}

\newtheorem{example}{Example}

\newcommand{\<}{\langle}
\renewcommand{\>}{\rangle}

\newcommand{\spand}{\hspace{0.5cm}\text{and}}

\DeclareMathOperator*{\argmax}{argmax}
\DeclareMathOperator*{\E}{E}
\DeclareMathOperator*{\mean}{mean}
\DeclareMathOperator*{\Var}{Var}

\newcommand{\rr}{\mathbb{R}}

\title{Computing  Black Scholes  with Uncertain Volatility-A Machine Learning Approach}

\author{ \href{https://orcid.org/0000-0002-5090-9218}{\includegraphics[scale=0.06]{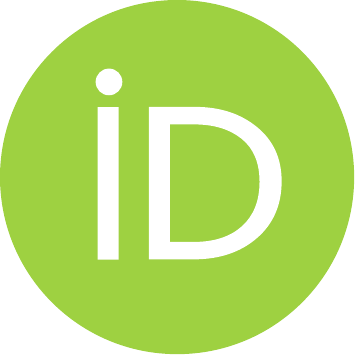}\hspace{1mm}Kathrin Hellmuth}\\
	Department of Mathematics\\
	University of W\"urzburg\\
	97074 Würzburg, Germany\\
	\texttt{kathrin.hellmuth@mathematik.uni-wuerzburg.de}\\
	\And 
	\href{https://orcid.org/0000-0003-2033-8204}{\includegraphics[scale=0.06]{orcid.pdf}\hspace{1mm}Christian Klingenberg} \\
	Department of Mathematics\\
	University of W\"urzburg\\
	97074 Würzburg, Germany\\
	\texttt{klingen@mathematik.uni-wuerzburg.de}\\
	}

\begin{document}
\maketitle
 \begin{abstract}
 In financial mathematics, it is a typical approach to approximate  financial  markets operating in discrete time by continuous-time models such as the Black Scholes model. Fitting this model gives rise to difficulties due to the discrete nature of market data. We thus model the pricing process of financial derivatives by the Black Scholes equation, where the volatility is  a function of a finite number of random variables. This reflects an influence of uncertain factors when determining volatility. The aim is to quantify the effect of this uncertainty when computing the price of derivatives. Our underlying method is the generalized Polynomial Chaos (gPC) method in order to numerically compute the uncertainty of the solution by the stochastic Galerkin approach and a finite difference method. We present an efficient numerical variation of this method, which is based on a machine learning technique, the so-called Bi-Fidelity approach. This is illustrated with numerical examples. 
 \end{abstract}

\newpage
\section{Introduction}
In modern financial markets, traders can choose from a large variety of financial derivatives. This term denotes financial instruments that have a value determined by so-called underlying variables or assets such as stocks, the oil price or the weather. Originally, derivatives were invented to reduce the risk of uncertain prices, especially in agricultural markets where one could have long periods between sowing and  harvest, see Chapter 1 and Chapter I in
\cite{WhaleyHistoryDerivatives, CrawfordThalesStory}, respectively.

As the derivative market was growing,  the need for a pricing formula for derivatives also increased in the 20th century. A breakthrough was made by  \cite{BSmodel} and \cite{Mertonmodel} when they contemporaneously formulated a model allowing the evaluation of derivatives. They were later awarded the Nobel prize in economics for their work. Derived from this model, the Black Scholes equation 
\begin{equation} \label{BSedet}
	\frac{\partial V(S,t)}{\partial t} + \frac{1}{2}\sigma^2 S^2\frac{\partial^2 V(S,t)}{\partial S^2} + rS\frac{\partial V(S,t)}{\partial S} - r V(S,t) = 0,  \hspace{0.5cm} S \in (0,\infty),\, t\in [0,T],
\end{equation}
explains the behaviour of the price $V$ of the derivative by means of a partial differential equation (PDE). This derivative is allowed to depend on the time $t$ up to maturity $T$ and only one underlying stochastic asset (e.g., a stock, an index or some commodity price). The price of the asset is denoted by $S$ and is assumed to follow a geometric Brownian motion. The constant $r$ denotes the risk free rate of interest in the market and $\sigma\in \mathbb{R}$ is the so-called volatility of the stochastic asset. 
Later, this model was extended to multiple underlying assets and adjusted for certain kinds of underlying variables, for example, interest rates; see \cite{CoxRossIngersollInterestRateDerivatives}.

Comparison to real data soon showed that the volatility $\sigma$ of one and the same stochastic asset can take values that differ more than explainable by rounding errors, etc.; see \cite{RubinsteinVolNotConst,ScottStochVol,Jungel}. The most popular approaches to deal with this are to model the volatility either as local volatility, i.e., a function $\sigma(S,t)$  as in  \cite{Dupire,ColemanEtAlLocalVol,CrepeyLocalVol,HankeRoslerLocalVol} or as a stochastic process; compare, e.g., the famous Heston model, \cite{HestonModel}, or \cite{RubinsteinVolNotConst,ScottStochVol,HullWhiteStochVol}.

These models as well as others are formulated in continuous-time. Because prices in financial markets are only visible in trades, the markets always operate in discrete time; compare, e.g.,  \cite{DiscreteTime_Mishura2021}. The models above, therefore, represent approximations of reality. In order to use them, they have to be fitted to the market using discrete data. This fitting procedure, however, introduces uncertainty in the recovered values, e.g., by rounding errors and by the interplay of the random nature of stochastic assets and their discrete observations. Because the interest rate $r$ is easy to determine even from discrete data, we focus on the volatility $\sigma$ for considerations of uncertainty.

Also in \cite{KoprivaSCollBS,PulchBSStochGal,DrakosBSStochGal}, the authors investigated uncertainty in the volatility:
They modelled it as a one-dimensional random variable $\Sigma(\omega)=\Theta(\omega)$  or a function of a one-dimensional random variable $\Sigma(\Theta(\omega))$ for $\omega$ in the underlying probability space. Then the price process $V(S,t,\Theta)$ also depends on $\Theta$  and follows the stochastic version of the Black Scholes equation
\begin{eqnarray} \label{BSestochVol}
	\frac{\partial V(S,t, \Theta)}{\partial t} + \frac{1}{2}\Sigma(\Theta)^2 S^2\frac{\partial^2 V(S,t, \Theta)}{\partial S^2} + rS\frac{\partial V(S,t, \Theta)}{\partial S} - r V(S,t, \Theta) = 0. 
\end{eqnarray}

This equation can be derived by inserting the  stochastic volatility into the Black Scholes model, where the Brownian motion is independent of $\Theta$. It can be studied by means of uncertainty quantification: 
The solution $V$ is developed in a generalized Polynomial Chaos (gPC) expansion
\begin{equation}\label{eq:IntroGPCV}
	V(S,t, \Theta(\omega)) = \sum_{n=0}^{\infty}v_n(S,t)p_n(\Theta(\omega))
\end{equation}
for orthonormal polynomials $p_n$ w.r.t. the distribution of $\Theta$ and coefficients given by the expected value $v_n(S,t) = E(V(S,t, \Theta)p_n(\Theta))$. If $\Theta$ has a density  $\mu:\mathcal{D}\to \mathbb{R}$, one can alternatively calculate the coefficients by
\begin{displaymath}
	v_n(S,t) = \int_{\mathcal{D}}V(S,t, x)p_n(x)\mu(x) \,dx.
\end{displaymath}

In \cite{KoprivaSCollBS}, these integrals are directly computed by a quadrature rule. 
The required  solutions $V(\cdot,\cdot,x_j)$ in the quadrature points $x_j$ 
are calculated as the solutions of the deterministic Black Scholes equation \eqref{BSedet} with $\sigma = x_j$.  This classifies the method as a Stochastic Collocation method.

In the articles  \cite{PulchBSStochGal,DrakosBSStochGal}, however, the stochastic Galerkin method is applied for computing the coefficients $v_n(S,t)$. By inserting the gPC expansion \eqref{eq:IntroGPCV} into the stochastic Black Scholes equation \eqref{BSestochVol}, multiplying the equation by an orthogonal polynomial $p_k(\Theta)$ and applying the expected value on both sides, deterministic PDEs for the coefficients $v_n(S,t)$  are derived
\begin{equation}\label{eq:IntroSGsystem}
	0= \frac{\partial v_k(S,t)}{\partial t} + \frac{1}{2}S^2 \sum_{n=0}^{\infty} \frac{\partial ^2 v_n(S,t)}{\partial S^2}E\left((\Sigma(\Theta))^2 p_k(\Theta)p_n(\Theta)\right)  + rS \frac{\partial  v_k(S,t)}{\partial S}  - r v_k(S,t).
\end{equation}

After truncating of the system and the coupling term to a finite number of indices, the system is solved numerically by the method of lines in \cite{PulchBSStochGal} and the finite element method in \cite{DrakosBSStochGal}.

In our work, we extend the model used above to the volatility $\Sigma(\Theta_1,\ldots,\Theta_L)$  depending on  many finitely random variables $\Theta_1,\ldots,\Theta_L$.
This leads to the stochastic Black Scholes equation 
\begin{eqnarray} \label{eq:BSestochVolLRV}
	0=&&\frac{\partial V(S,t,\Theta_1,\ldots,\Theta_L )}{\partial t} + \frac{1}{2}\Sigma^2(\Theta_1,\ldots,\Theta_L) S^2\frac{\partial^2 V(S,t,\Theta_1,\ldots,\Theta_L)}{\partial S^2} \\ &&+rS\frac{\partial V(S,t, \Theta_1,\ldots,\Theta_L)}{\partial S} - r V(S,t, \Theta_1,\ldots,\Theta_L). \nonumber 
\end{eqnarray} 

A model like this might, for instance, occur when the volatility is modelled as a random variable that also depends on certain stochastic factors as in 
{\cite{zhang2018policy, BAZZANA2020101240, xie2021does, laws10030055}}.  
We propose a momentum constrained maximum likelihood technique to fit the volatility distribution to real data and compare our results to market data. This enables the comparison of our model to real data, which was not considered in literature yet, but is an important tool to measure the model fit.

The solution is then derived in the same way as in \eqref{eq:IntroSGsystem} and calculated numerically by a finite difference method. The increased computational cost for multiple similar calculations is reduced by a Bi-Fidelity technique, which can be considered as a machine learning approach, whose effectiveness is illustrated in numerical examples.

The outline of the article is as follows: After introducing gPC to finitely many random variables,  a method of fitting the stochastic volatility to real data is described in Section \ref{sec:gPCFitVol}. The stochastic Galerkin method is used to solve Equation \eqref{eq:BSestochVolLRV} in section \ref{sec:SG}. However, computational costs can be  high. Thus, we introduce a Bi-Fidelity numerical technique to compute this more efficiently in Section \ref{sec:BiFid}. The paper is rounded out with numerical results illustrating the effectiveness of this technique and the fit to real market data in Section \ref{sec:Results}.

\section{Fitting the Random Volatility to Real Market Data}\label{sec:gPCFitVol}

For the convenience of the reader and in order to introduce notation, we briefly recall the fundamentals of generalized polynomial chaos (gPC). We then propose a method to fit the volatility distribution to real data.

Denote by $\Theta_1,\ldots,\Theta_L$ random variables with joint distribution function $\bar{F}:\bar{\mathcal{D}} \to \mathbb{R}$ for a multivariate domain $\bar{\mathcal{D}}\subset \mathbb{R}^L$. For a function $\bar{f}:\bar{\mathcal{D}} \to \mathbb{R}$, the following notation is used for integration with respect to (w.r.t.) $\bar{F}$:
\begin{displaymath}
	\< \bar{f}\> := \int_{\bar{\mathcal{D}}} \bar{f}(x_1,\ldots,x_L)\,d\bar{F}(x_1,\ldots,x_L) = E(\bar{f}(\Theta_1,\ldots,\Theta_L)).
\end{displaymath}

Consider a system of polynomials $\{\bar{p}_{\alpha}:\bar{\mathcal{D}}\to \mathbb{R}\,|\,\alpha= (\alpha_1,..,\alpha_L)\in \mathbb{N}_0^L\}$, where the polynomial $\bar{p}_{\alpha}(x_1,\ldots,x_L)$ has degree in the $i$-th variable $deg_{x_i}(\bar{p}_{\alpha})= \alpha_i$. In adaption to Definition 8.24 in \cite{SullivanUQ}, we call this an \textit{infinite system of orthogonal polynomials w.r.t. $\bar{F}$}, if for all multi indices $\alpha, \beta \in \mathbb{N}_0^L$ one has
\begin{eqnarray*}
	&&\< \bar{p}_{\alpha} \bar{p}_{\beta}\> = 0 \hspace{0.9cm} \text{for }\alpha \neq \beta,\\
	&&\< \bar{p}_{\alpha}^2\> =: \bar{\gamma}_{\alpha}>0. 
\end{eqnarray*}

The existence of orthogonal polynomial systems follows from the Gram Schmidt algorithm, if for all $\alpha=(\alpha_1,\ldots,\alpha_L)\in \mathbb{N}_0^L$ the moments $\<x_1^{\alpha_1}\cdot\ldots\cdot x_L^{\alpha_L}\>$ are finite. Uniqueness of the orthogonal polynomials is then given up to multiplication by constants. If the $\Theta_i$ are independent, they are in particular given by the product of the orthogonal polynomials w.r.t. the distribution of each $\Theta_i$.

In the following, $L^p_{dF}(D,H)$ denotes the space of all functions $D \to H$ that are $p$-times integrable w.r.t. the measure $dF$ for some $D\subset \mathbb{R}^n$ and codomain $H$. If $dF$ is not explicitly defined, the Lebesgue measure is chosen. If $D$ and $H$ are not defined, then $D$ equals the domain of $F$ and $H$ equals $\mathbb{R}$. 

It is well known  that under certain circumstances, orthogonal polynomials span the space $L^2_{d\bar{F}}$. They are thus called a complete orthogonal basis of  $L^2_{d\bar{F}}$.

This is, for example, the case, if $\bar{F}$ is absolutely continuous, has finite moments and either $(\Theta_1,\ldots,\Theta_L)$ realizes in a compact domain almost surely or the density of  $\bar{F}$ is exponentially integrable. For details{, see  \cite{RahmangPCDependency}}. If the $\Theta_i$ are independent, the orthogonal polynomials w.r.t. $\bar{F}$ span $L^2_{d\bar{F}}$, if all orthogonal polynomial systems w.r.t. the density of $\Theta_i$ span the corresponding $L^2$ spaces. This is due to the tensor product representation of $L^2_{d\bar{F}}$ in case of independency of the $\Theta_i$, see Example E.10 in \cite{JansonLpIsom}.

Assuming such circumstances to be given, the gPC expansion can be defined. 
\begin{theorem}[adaption of section 11.3 
	in \cite{SullivanUQ}]\label{th:gPCmultivar}
	Let $\Theta_1,\ldots,\Theta_L:\Omega\to \mathbb{R}$ be random variables with joint distribution $\bar{F}$ such that the orthogonal polynomials $(\bar{p}_{\alpha})_{\alpha\in\mathbb{N}_0^L}$ w.r.t. $\bar{F}$ form a complete basis of $L^2_{d\bar{F}}$. Denote by $\mathcal{H}$ an arbitrary Hilbert space, e.g., the real numbers $\mathbb{R}$ or a space of the form $ L^p(D,\mathbb{R})$, $p=0,1,2$, for some domain $D\subset \mathbb{R}^n$. Then every random variable  $X:\Omega \to \mathcal{H}$ with 
	\begin{equation}\label{eq:formX}
		X=^d \tilde{X}(\Theta_1,\ldots,\Theta_L)
	\end{equation}
	in distribution for a function $\tilde{X}\in L^2_{d\bar{F}}(\bar{\mathcal{D}}, \mathcal{H})$  
	can be represented in the generalized Polynomial Chaos {(gPC)} form 
	\begin{equation}\label{gPCmultivar}
		X=^d \sum_{\alpha \in \mathbb{N}_0^L}x_{\alpha} \bar{p}_{\alpha}(\Theta_1,\ldots,\Theta_L) \hspace{1cm}\text{with} \hspace{1cm} x_{\alpha} = \frac{\<X\bar{p}_{\alpha}\>}{\<\bar{p}_{\alpha}^2\>}\in \mathcal{H}.
	\end{equation} 
\end{theorem}

The proof follows in analogy to the proof for independent continuous random variables in Section 11.3 in \cite{SullivanUQ}  from the tensor product decomposition $ L^2_{d\bar{F}}\otimes \mathcal{H}\cong L^2_{d\bar{F}}(\bar{\mathcal{D}}, \mathcal{H})$.


Assuming  $\Sigma \in L^2_{d\bar{F}}$,  Theorem \ref{th:gPCmultivar} gives the decomposition of the volatility 
\begin{align}
	\Sigma(\Theta_1,\ldots,\Theta_L) := \sum_{\alpha \in \mathbb{N}_0^L} \sigma_{\alpha} \bar{p}_{\alpha}(\Theta_1,\ldots,\Theta_L){.}\label{eq:gPCSigmamultivar}
\end{align}

To fit the model to the data, we truncate the series by bounding the maximum polynomial degree $|\alpha|:= \alpha_1+\ldots+\alpha_L$  by $K\in \mathbb{N}_0$
\begin{equation}
	\Sigma^K(\Theta_1,\ldots,\Theta_L) := \sum_{\alpha \in \mathbb{N}_0^L, \, |\alpha|\leq K } \sigma_{\alpha} \bar{p}_{\alpha}(\Theta_1,\ldots,\Theta_L).\label{eq:truncatedgPCSigmamultivar}
\end{equation}

We propose a momentum constrained maximum likelihood approach to fit the gPC coefficients $\sigma_{\alpha}$ to discrete real-world data. This facilitates the computation.

The values of the volatility of an asset can be obtained, e.g., by calculating the implied volatilities from corresponding European options. This generates a dataset of implied volatilities representing observations of the random variable $\Sigma^K(\Theta_1,\ldots,\Theta_L)$. When fitting the volatility to the data, we constrain ourselves to those tuples of coefficients corresponding to a volatility $\Sigma^K(\Theta_1,\ldots,\Theta_L)$ whose first  moments coincide with the empirical moments of the dataset. The choice of {these constraints reduces}  the dimension of {the parameter space for} the likelihood function { while  important characteristics of the distribution---the statistical moments---are maintained.}

We illustrate the technique on the simple case of two independent random variables and truncation at $K=1$:
\begin{example}\label{ex:getSigma}
	Let  {$\Theta_1,\Theta_2$} be two independent random variables of known distribution and w.l.o.g. expected value $0$ and variance $1$. {Assume} the densities of their distributions exist and {denote them} by $f_1,f_2$, respectively. Now consider the  random volatility truncated to maximum polynomial degree~$1$
	\begin{displaymath}
		\Sigma^1(\Theta_1,\Theta_2) = \sigma_{00} \bar{p}_{00}(\Theta_1,\Theta_2) + \sigma_{01} \bar{p}_{01}(\Theta_1,\Theta_2) + \sigma_{10} \bar{p}_{10}(\Theta_1,\Theta_2) = \sigma_{00}  + \sigma_{01} \Theta_2 + \sigma_{10} \Theta_1
	\end{displaymath}
	when choosing orthonormal polynomials. Assume that values of the volatility $y_1,\ldots,y_M$ are given. \\
	The plain maximum likelihood method would maximize the joint density $h$ of the realizations  $y_1,\ldots,y_M$ of $\Sigma^1(\Theta_1,\Theta_2)$ w.r.t. the three coefficients $(\sigma_{00}, \sigma_{01}, \sigma_{10})\in U\subset\rr^3\setminus\{0\}$: 
	\begin{align*}
		(\sigma_{00}, \sigma_{01}, \sigma_{10})& = \argmax_{(\sigma_{00}, \sigma_{01}, \sigma_{10})\in U} h(y_1,\ldots,y_M\mid (\sigma_{00}, \sigma_{01}, \sigma_{10})) \\
		&=  \argmax_{(\sigma_{00}, \sigma_{01}, \sigma_{10})\in U} \prod_{i=1}^M \frac{1}{|\sigma_{10}\sigma_{01}|}\int_{t\in \rr}f_1\left(\frac{x-\sigma_{00}-t}{\sigma_{10}}\right)f_2\left(\frac{t}{\sigma_{01}}\right)dt.
	\end{align*}
	
	Constraining the maximum likelihood estimator to be exact in the first moment, i.e., the expected value, gives
	\begin{displaymath}
		\E(\Sigma^1(\Theta_1,\Theta_2)) = \sigma_{00} \overset{\text{!}}= \bar{y}_M:=\mean(y_1,\ldots,y_M).
	\end{displaymath}
	
	This reduces the optimization to two variables 
	\begin{displaymath}
		(\sigma_{01}, \sigma_{10}) = \argmax_{( \sigma_{01}, \sigma_{10})\in \hat{U}}  \prod_{i=1}^M \frac{1}{|\sigma_{10}\sigma_{01}|}\int_{t\in \rr}f_1\left(\frac{x-\bar{y}_M-t}{\sigma_{10}}\right)f_2\left(\frac{t}{\sigma_{01}}\right)dt,
	\end{displaymath}
	where $\hat{U}= U\cap \{\bar{y}_M\}\times \rr^2$. 
	
	Further constraints can be used to reduce the complexity even more: Claiming, e.g., the variance to coincide with the empirical variance $\Var(\Sigma^1(\Theta_1,\Theta_2))\overset{\text{!}}= S_M^2$ gives the additional relation 
	\begin{displaymath}
		\sigma_{10} = \pm\sqrt{S_M^2-\sigma_{10}^2}
	\end{displaymath}
	reducing the optimization to one variable and the sign of  $\sigma_{10}$. This ends our example.
\end{example}
%
\section{Deriving the System of PDEs for the gPC Coefficients}\label{sec:SG}
The stochastic Galerkin method is applied to the Black Scholes equation \eqref{eq:BSestochVolLRV} with uncertain volatility
in order to transform the stochastic PDE into a system of deterministic PDEs for the gPC coefficients of the solution $V(S,t,\Theta_1,\ldots,\Theta_L )$. 

To do so, one has to assume  $\Sigma \in L^2_{d\bar{F}}$ and $V\in  L^2_{d\bar{F}}(\bar{\mathcal{D}}, L^2((0,\infty)\times[0,T], \mathbb{R}))$, such that Theorem \ref{th:gPCmultivar} can be applied. 
In analogy to the one-dimensional case in \cite{DrakosBSStochGal,PulchBSStochGal}, the thus derived gPC expansions are inserted in the Black Scholes equation \eqref{eq:BSestochVolLRV}. Multiplication of the equation by $\bar{p}_{\delta}(\Theta_1,\ldots,\Theta_L)$ and application of the expected value, for each $\delta\in \mathbb{N}_0^L$ at a time, yields the equations
\begin{equation*}
	0 = \frac{\partial v_{\delta}(S,t)}{\partial t}
	+ \frac{1}{2}S^2\sum_{\alpha, \beta, \gamma\in \mathbb{N}_0^L} \sigma_{\alpha}\sigma_{\beta}\frac{\partial^2 v_{\gamma}(S,t)}{\partial S^2} M_{\alpha \beta \gamma \delta}
	+ rS \frac{\partial v_{\delta}(S,t)}{\partial S} - r  v_{\delta}(S,t)
\end{equation*}
due to orthogonality of the $p_{\alpha}$. Note that the Galerkin multiplication tensor $M_{\alpha\beta\gamma\delta}:=\frac{\<\bar{p}_{\alpha}\bar{p}_{\beta} \bar{p}_{\gamma}\bar{p}_{\delta}\>}{\<\bar{p}_{\delta}^2\>}$  exists since the integrated functions are all polynomials in $L$ variables.

In order to solve the system, the boundary conditions and the  final condition corresponding to the considered financial derivative  are transformed to conditions on the gPC coefficients $v_i$.  Usually, they are deterministic and thus appear in the coefficient $v_{(0,\ldots,0)}$, whereas all other coefficients vanish.

We  truncated the gPC expansions up to maximum polynomial degrees $K,N\in \mathbb{N}_0$ and obtained representation \eqref{eq:truncatedgPCSigmamultivar} for $\Sigma^K$ and 
\begin{alignat}{4}
	V^N(S,t, \Theta_1,\ldots,\Theta_L) &:=&&\sum_{\delta\in \mathbb{N}_0^L, \, |\delta|\leq N} v^N_{\delta}(S,t)\bar{p}_{\delta}(\Theta_1,\ldots,\Theta_L) \label{eq:truncatedgPCVmultivar}
\end{alignat}
for  coefficients $ v^N_{\delta} \in L^2((0,\infty)\times [0,T], \mathbb{R})$. 

The system of equations for the truncated gPC coefficients $v^N_{\delta}$,  $\delta \in \mathbb{N}_0^L$ with $|\delta|\leq N$,  is then given by
\begin{equation}\label{eq:truncateSGsystemMultivar}
	0 = \frac{\partial v^N_{\delta}(S,t)}{\partial t}
	+ \frac{1}{2}S^2\sum_{\substack{\alpha, \beta, \gamma\in \mathbb{N}_0^L, \\ |\alpha|,|\beta|\leq K, \\ |\gamma| \leq N}} \sigma_{\alpha}\sigma_{\beta}\frac{\partial^2 v^N_{\gamma}(S,t)}{\partial S^2} M_{\alpha\beta\gamma\delta}
	+ rS \frac{\partial v^N_{\delta}(S,t)}{\partial S} - r  v^N_{\delta}(S,t),
\end{equation}
which can be evaluated numerically. For demonstrative purposes, we use a finite difference scheme, see Appendix \ref{sec:appendix}.

Note, however, that convergence of the truncated stochastic Galerkin solution $V^N$ in~\eqref{eq:truncatedgPCVmultivar} to the true solution $V$ as $N\to \infty$ is not obvious and could not be proven to date. It is a topic open to further research. From now on, we assume convergence.

\section{A Bi-Fidelity Approach for Calculating the Stochastic Galerkin Solution to the Black Scholes Equation with Random Volatility}\label{sec:BiFid}
If the volatility depends on just $L=2$ random variables, the stochastic Galerkin (SG) solution truncated at maximum degree $N$ already has $(N+1)(N+2)/2$ gPC coefficients. Thus, $(N+1)(N+2)/2$ coupled equations have to be solved to obtain the approximate SG solution. The number of equations and with it the computational cost rapidly increase as $N$ or $L$ increases. 

This becomes a problem especially if the SG solutions for many options shall be computed at a time, e.g., for risk management evaluations of derivatives with different underlying assets. The Bi-Fidelity approach provides a solution to this problem, if the same type of option (e.g., European Call options) with the same maturity $T$ and interest rate $r$, but different distributions of the volatility model $\Sigma(\Theta_1,\ldots, \Theta_L)$ are considered. A situation like this arises, for instance, when comparing financial derivatives of the same type and maturity date but with different underlying stochastic assets.

{In general, g}iven a PDE depending on a random variable $\Xi$, the Bi-Fidelity method aims  to approximate the desired high fidelity solution at a certain realization $z$ of $\Xi$ by pre-stored high and low fidelity solutions in some other realizations of $\Xi$ and the computationally cheaper low fidelity solution in $z$. This can be considered a machine learning approach because the properties of the equation are learned offline by picking suitable  realizations for the stored approximation data.

{The application of  Bi-Fidelity techniques to various problems is an active area of research, see e.g. \cite{de2020bi, FAIRBANKS2020108996}. In the setting of uncertainty quantification for PDE models, it is frequently described in the context of uncertainty quantification via Stochastic Collocation methods, compare \cite{ZhuNarayanXiuStochCollBiFid,NarayanGittelsonXiuStochCollBiFid} for the general procedure or \cite{LiuBiFid, Gamba_2021, GAO2020113047, liu2021bi} for {applications}. {The combination with the stochastic Galerkin method works similarly; however, it is not very common in literature.}}

At first, the random variable $\Xi$ has to be assigned. In our case, it is not given by $(\Theta_1,\ldots,\Theta_L)$, since we still want our solution to be a random variable depending on the $\Theta_i$ in order to explore its stochastic behaviour. Instead, we suppose the distribution of  $\Sigma(\Theta_1,\ldots,\Theta_L)$ to change from calculation to calculation, as it would be the case for different underlying assets, without changes in the distributions of the $\Theta_i$. This could reflect different sensitivities of different stochastic assets to the factors $\Theta_i$. By representation~\eqref{eq:truncatedgPCSigmamultivar} of the truncated gPC expansion of $\Sigma$, a variation in the distribution of the volatility  means a variation in at least one of the gPC coefficients $\sigma_{\alpha}$, $|\alpha|\leq K$. Hence, the random variable $\Xi$ describes volatility models of the form \eqref{eq:truncatedgPCSigmamultivar} by their gPC coefficients $\sigma_{\alpha}, |\alpha|\leq K$.

Then, high and low  fidelity models have to be defined. The high fidelity model is the one we are actually interested in. We choose a high-resolution numerical solution to \eqref{eq:truncateSGsystemMultivar} 
derived by the explicit finite difference scheme \eqref{eq:fdScheme}
with a large number of grid points in the $S$-space  $M^H_{\zeta}$ and in time $N^H_{\tau}$.  The low fidelity model, i.e., the cheaper model that is less trusted but used for the approximation rule, is chosen to be the same numerical solution on a coarse grid with small $M^L_{\zeta}$ and $N^L_{\tau}$.

Note, however, that $N_{\tau}^L$ must not be chosen too small; otherwise, the stability of the scheme could be violated for a large number of volatility models. The reason for this requirement will become clear at step 1 of the  offline data generating steps. 

It is also possible to consider  different numerical schemes or even different but similar underlying equations for the high and low fidelity models. However, important characteristics of the solution must be shared between the models.

Now one can proceed with the typical Bi-Fidelity algorithm as described in \cite{NarayanGittelsonXiuStochCollBiFid,ZhuNarayanXiuStochCollBiFid,LiuBiFid}.
Below, the application of this algorithm is explained, where the volatility is assumed to depend on $L=2$ random variables $\Theta_1, \Theta_2$ for a better readability. An extension to more random variables can easily be achieved. The truncation number $K=1$ is chosen such that the random variable $\Xi$ represents the gPC coefficients $\sigma_{00}, \sigma_{10}$ and $\sigma_{01}$ of the volatility as in Example \ref{ex:getSigma}.

Since the actual computational effort lies in the calculation of the transformed system of Equation \eqref{eq:vectortranftruncSG},
the Bi-Fidelity approach is applied directly on the transformed $\mathbf{\bar{v}}$. Thus, a transformation back to the original variables $\mathbf{v}^N, S$ and $t$ is  performed only once for the Bi-Fidelity solution, reducing the computational cost. For the calculation of the scheme, initial conditions and the Galerkin multiplication tensors are pre-stored and reused.

The following three steps describe the \textit{offline learning phase} of the algorithm in which the stored approximation data are generated, compare e.g. \cite{NarayanGittelsonXiuStochCollBiFid,ZhuNarayanXiuStochCollBiFid,LiuBiFid}. These steps have to be executed only once.
\begin{enumerate}[label= Step \arabic*:]
	\item \label{step:BiFid1manyLowFid}
At first, the codomain of $\Xi$ is described by finite intervals $\sigma_{00} \in [a_{00}, b_{00}], \sigma_{10}\in [a_{10},b_{10}], \sigma_{01} \in [a_{01}, b_{01}]$.

The intervals can,  for instance, be constructed  by  calculation of  $\sigma_{00}, \sigma_{10}, \sigma_{01}$ for some of the  stochastic assets of interest. 
Alternatively, one can think of possible values of $\sigma_{00}$ inspired by similar experiments and choose bounds of $\sigma_{10}$ and $\sigma_{01}$ such that the variance of $\Sigma(\Theta_1, \Theta_2)$ is bounded by some predefined value. We used this approach in the calculations of Section \ref{sec:Results}.

After that, a large set $Y$ of possible realizations of $\Xi$ has to be chosen such that it is a good 'cover' of the possible values of $\Xi$. One can use Monte Carlo sampling or a structured grid on the codomain of $\Xi$. 

For every volatility model described by a $y\in Y$, the low fidelity solution $\mathbf{\bar{v}}^{L}(y)$ is computed, if the corresponding system of equations is parabolic and the low fidelity scheme is stable.
\item Since one can usually not afford to calculate the high fidelity solution in every $y\in Y$, one has to determine the most important points. Let  $A\in \mathbb{N}$ denote the number of high fidelity computations one can afford, then this can be achieved by choosing $z_0 := \argmax_{y\in Y} d^L(\mathbf{\bar{v}}^{L}(y), {0}))$ and
\begin{equation}\label{eq:BiFidPointSelection}
	z_{i+1}:= \argmax_{y\in Y} d^L(\mathbf{\bar{v}}^{L}(y), \mathbf{\bar{V}}^{L}({z_1,\ldots,z_i})), \hspace{0.5cm} i=0,\ldots,A-1.
\end{equation}

The notation $\mathbf{\bar{V}}^L(\hat{Y}):= \text{span}(\mathbf{\bar{v}}^{L}(\hat{y})\,|\, \hat{y}\in \hat{Y})$ for $\hat{Y} \subset Y$ is used for the span of the solutions to previously picked $z_i\in \hat{Y}$. Then $d^L(u, V):= \inf_{v\in V}\|u-v\|^L$ is the distance  of a point $v\in \mathbf{\bar{V}}^L(Y)$ to the set $V\subset \mathbf{\bar{V}}^{L}(Y)$.  We used a greedy procedure for the point selection; for further details on the computation, compare Algorithm 1 in \cite{NarayanGittelsonXiuStochCollBiFid}. 

This step selects the points $z_1,\ldots,z_A$ whose solutions span the 'largest' subspace $\mathbf{\bar{V}}^{L}({z_1,\ldots,z_A})$ of $\mathbf{\bar{V}}^{L}(Y)$.
\item The high fidelity solution is calculated in the thus derived points $z_1,\ldots,z_A$. Note that $N^H_{\tau}$ has to be chosen large enough such that the numerical scheme is stable for all volatility models $z_i$. The parabolicity of the system of PDEs does not need to be checked again, as it has been checked  in step 1 already. The pairs of high and low fidelity solutions $\mathbf{\bar{v}}^{H}(z_i), \mathbf{\bar{v}}^{L}(z_i)$ are stored.
\end{enumerate}

Assume now that a certain volatility model $z$ is given and  the corresponding Bi-Fidelity solution of the Black Scholes equation with uncertain volatility shall be computed. This is performed in the \textit{online phase} as follows, see  \cite{NarayanGittelsonXiuStochCollBiFid,ZhuNarayanXiuStochCollBiFid,LiuBiFid}:
\begin{enumerate}[label= Step \arabic*:]
\item 
The low fidelity solution $\mathbf{\bar{v}}^L(z)$ is calculated by scheme \eqref{eq:fdScheme}. Note that the system of equations needs to be parabolic and the scheme has to be stable for a reasonable calculation.

\item 
The low fidelity solution $\mathbf{\bar{v}}^L(z)$ is projected onto $\mathbf{\bar{V}}^{L}({z_1,\ldots,z_A})$ leading to the projection formula
\begin{displaymath}
\bar{\mathbf{v}}^L(z) \approx P_{\mathbf{\bar{V}}^L({z_1,\ldots,z_A})}\mathbf{\bar{v}}^L(z) = \sum_{k=1}^A c_k \mathbf{\bar{v}}^L(z_k)
\end{displaymath} 
with projection coefficients $c_k\in \mathbb{R}$. Here $P_{\mathbf{V}}\mathbf{y}$ denotes the orthogonal projection of $\mathbf{y}$ onto $\mathbf{V}$. Details of the computation of the $c_k$ can be found in \cite{NarayanGittelsonXiuStochCollBiFid}, for instance.

\item 
Finally, the Bi-Fidelity solution is constructed by applying the same projection law to the stored high fidelity solutions  
\begin{equation*}
\mathbf{\bar{v}}^{BF}(z):= \sum_{k=1}^A c_k \mathbf{\bar{v}}^H(z_k).
\end{equation*}
\end{enumerate}
After deriving $\mathbf{\bar{v}}^{BF}$, it has to be transformed back to the original variables $\mathbf{v}$, $S$ and $t$.

\section{Numerical Results}\label{sec:Results}
This section presents numerical solutions to the Black Scholes equation with uncertain volatility. For the sake of simplicity, the volatility is assumed to depend on the two independent random variables $\Theta$ with standard normal distribution and $\Delta$  uniformly distributed on $[-0.5,0.5]$. The properties of such models are investigated and the model is tested on real data. Furthermore, the error of the Bi-Fidelity approximation is investigated and its computation  time  is compared to the high fidelity model.

For more convenient reading, times $t$ and the maturity $T$ are given in days, whereas for the computations, these values were multiplied by $1/251$ to go over to years.
\subsection{Results for the Extended Model}
\begin{figure}[H]
\centering
\begin{subfigure}{8cm}
	\centering
	\includegraphics[width = 8cm]{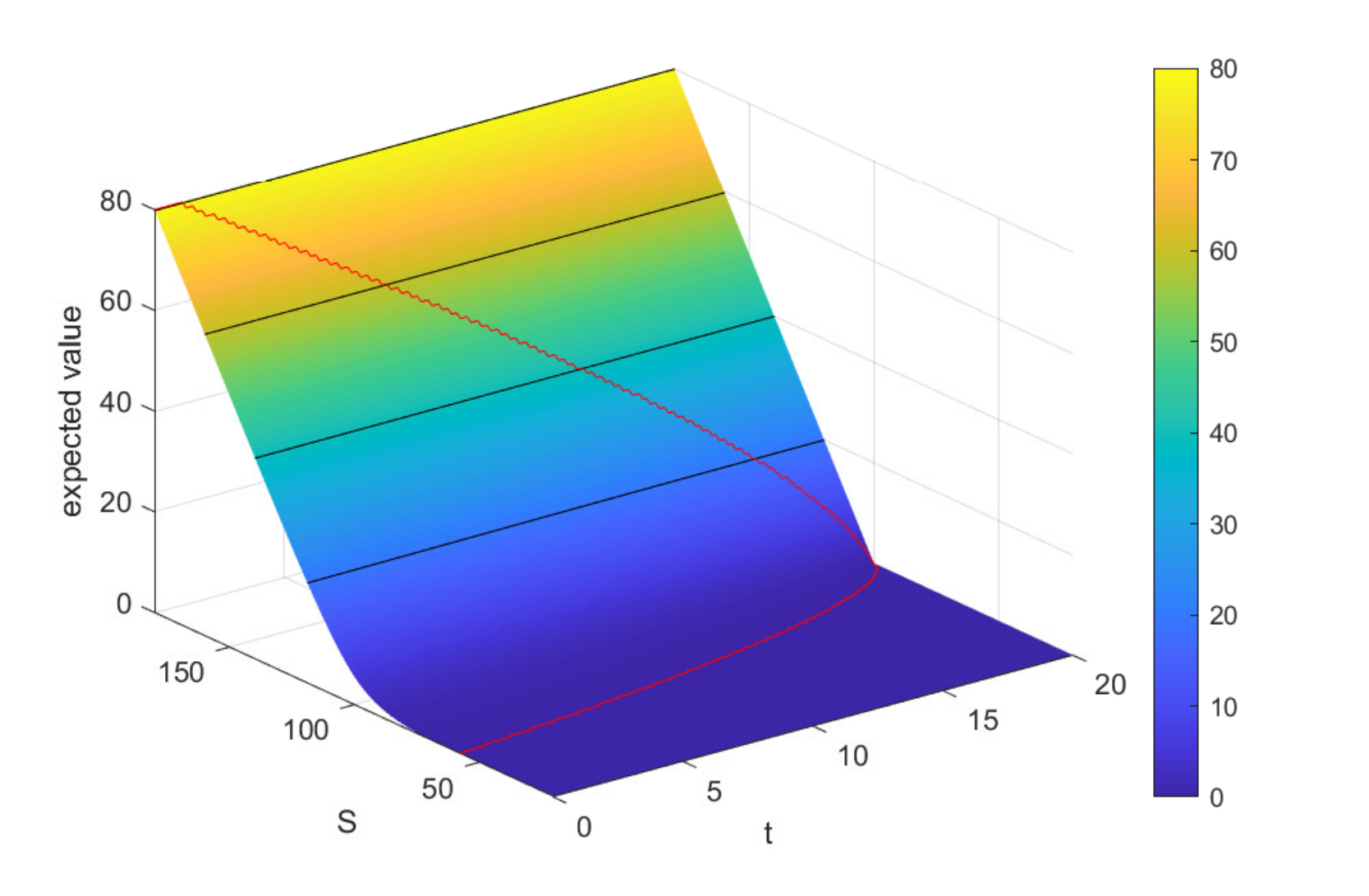}
	\caption{Expected value surface for the stochastic solution.}\label{fig:HermLegMean}
\end{subfigure}~
\begin{subfigure}{8cm}
	\centering
	\includegraphics[width = 8cm]{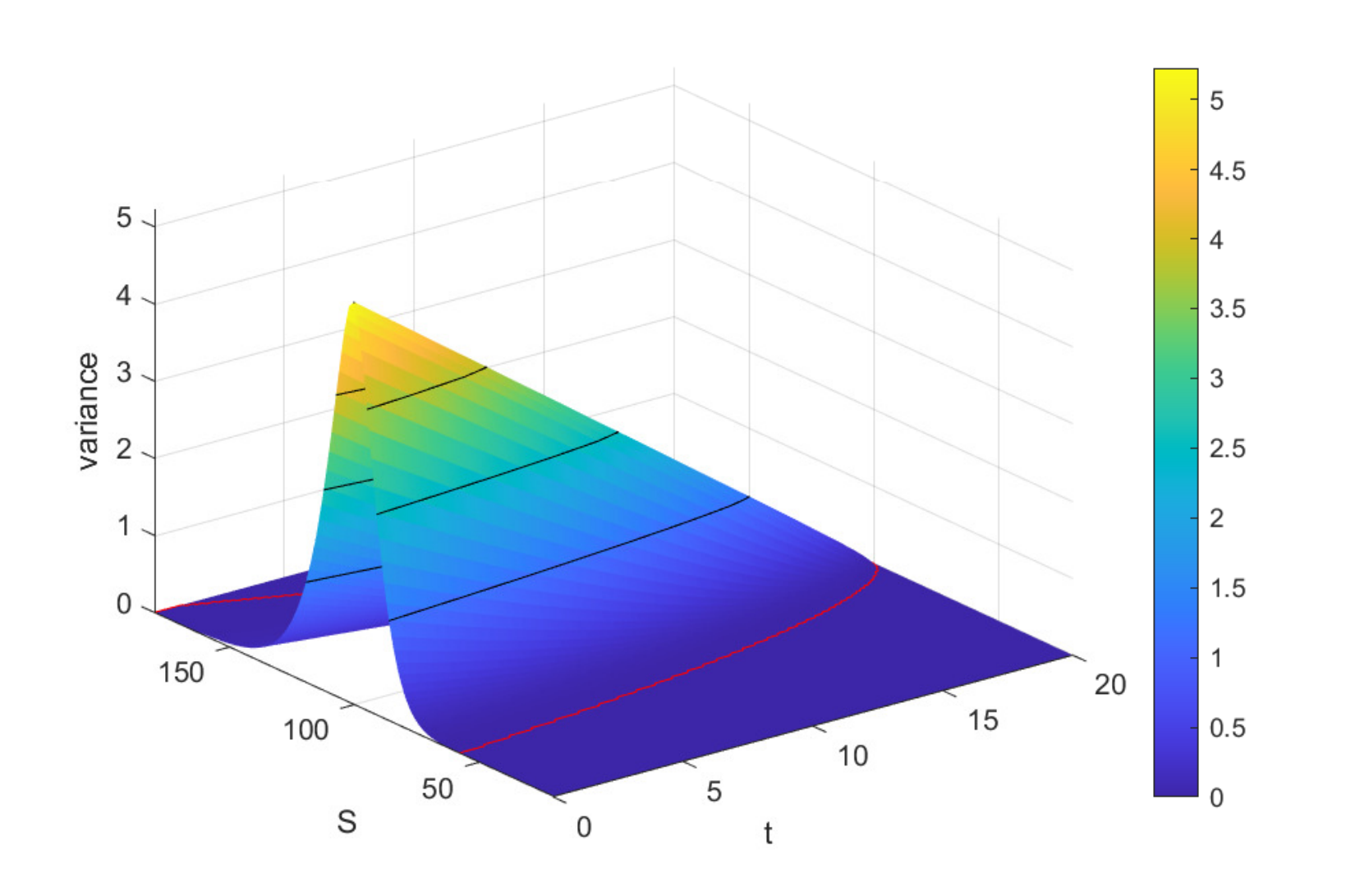}
	\caption{Variance surface for the stochastic solution.}\label{fig:HermLegVar}
\end{subfigure}
\\	
\begin{subfigure}{8cm}
	\centering
	\includegraphics[width = 8cm]{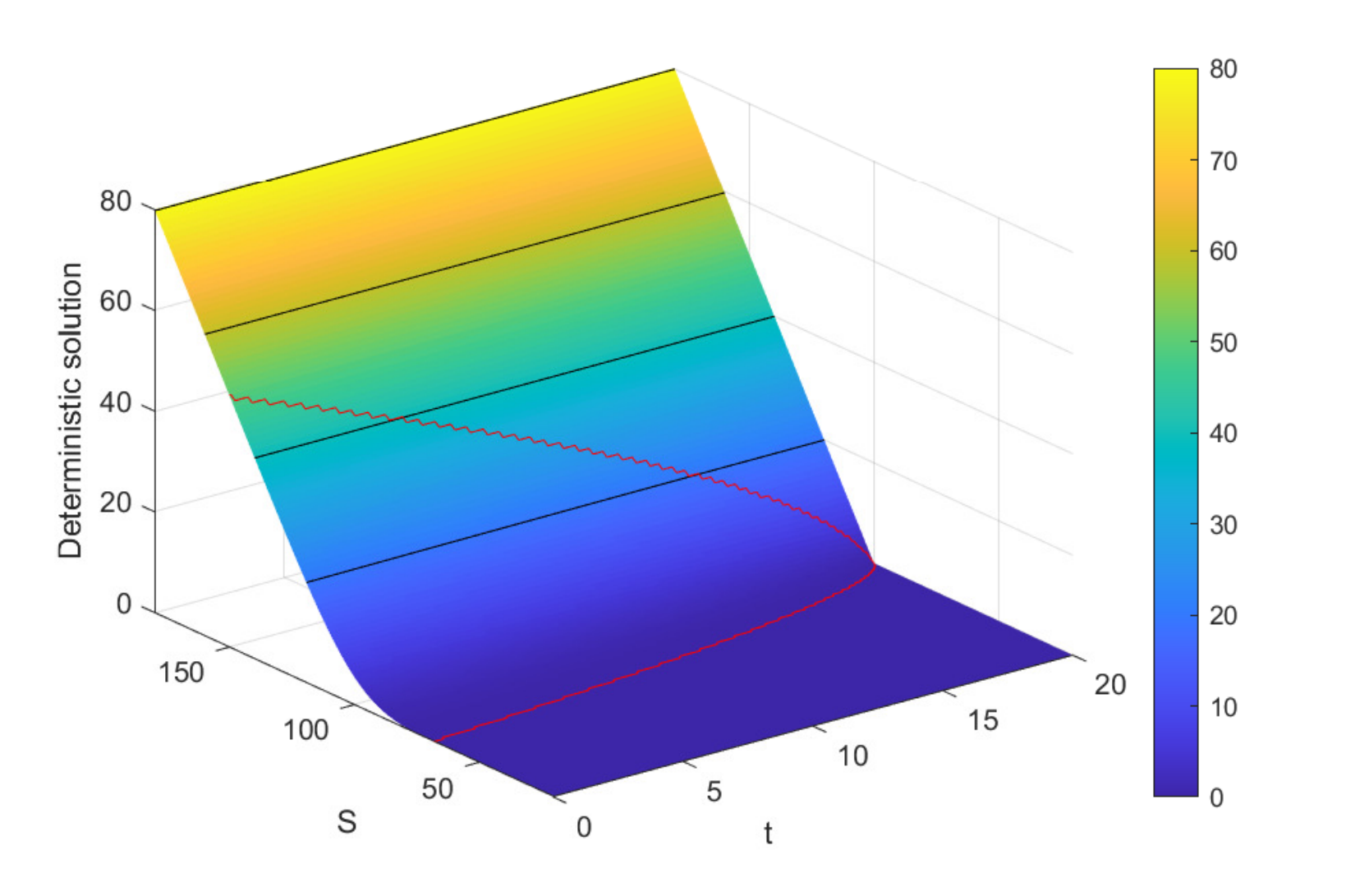}
	\caption{Deterministic solution.}\label{fig:Det}
\end{subfigure}
\caption{Solutions to the Black Scholes equation for a European Call option with $T=20, strike =100$ and $r=0$ for the volatility model $\Sigma_1(\Theta, \Delta) = 0.5+0.2\Theta +0.1\sqrt{12}\Delta$, $\Theta$ normally and $\Delta$ uniformly distributed, in (\textbf{a},\textbf{b}) and the deterministic model $\sigma=0.5$ in (\textbf{c}) calculated with $K=1,N=5$, $M_{\zeta}=200$, $N_{\tau} = 319$.}
\label{fig:MeanVarDet}
\end{figure}

The numerical solution to the truncated system of Equation \eqref{eq:truncateSGsystemMultivar} for a European Call option with a strike price $strike=100$ and maturity $T=20$ in a market with risk-free rate of interest $r=0$ is visualized in Figure \ref{fig:MeanVarDet}a,b by plotting its mean and variance. 

The volatility of the underlying stochastic asset is modelled by
\begin{equation}\label{eq:volModel}
\Sigma_1(\Theta, \Delta) = 0.5+0.2\Theta +0.1\sqrt{12}\Delta
\end{equation}
for  $\Theta$ standard normally distributed and  $\Delta$ uniformly distributed on $[0.5,0.5]$. The random variables are modelled to be independent.
For the gPC expansion of the solution, the truncation number $N=5$ was chosen, for which system \eqref{eq:truncateSGsystemMultivar} is parabolic. The numbers of grid points $M_{\zeta}=200$ in $\zeta$ and $N_{\tau}=319$ in $\tau$ were chosen such that the applied explicit finite difference scheme \eqref{eq:fdScheme} is stable.

Contour lines were drawn at a height of quarters of the maximum absolute value and the borders of the smoothing area, i.e., the area where the solution differs from its final condition $V(S,T) = (S-strike)^+$, were drawn in red. These lines will be present in each of the following surface plots. Note that the expected value surface resembles the solution of the deterministic Black Scholes equation for $\sigma = 0.5$ in Figure \ref{fig:MeanVarDet}c, but the smoothing area is larger.

Experiments showed that the qualitative shape of the expected value and variance is characteristic for solutions to the Black Scholes equation with random volatility \eqref{eq:BSestochVolLRV}  of the form $\Sigma(\Theta, \Delta)=\sigma_{00}+\sigma_{10}\Theta +\sigma_{01}\Delta$. These models lead to solutions that 'lie between' the solutions for volatility that depends on $\Theta$ or $\Delta$ only   and has the same mean and variance. 

The higher $\sigma_{10}$ is in comparison to $\sigma_{01}$, the closer the solution is to the solution for volatility depending on $\Theta$ only and the further away it is from the solution for the model depending on $\Delta$ only, and vice versa. An increase in the mean $\sigma_{00}$ of the volatility while keeping its variance constant was observed to enlarge the smoothing area and thus the spread of the variance, which in turn flattens it. 

A rise in the variance $\sigma_{10}^2+\sigma_{01}^2/12$ of the volatility with constant mean $\sigma_{00}$, however, seemed to scale up the variance of the SG solution by the same factor. Meanwhile, the expected value of the SG solution was marginally increased within the smoothing area. 
\newpage
\subsection{Comparison to Real Market Data}

The model is compared to market prices of a European Call option, whose end of day values are considered from 7 January 2019 to 20 September 2019. \footnote{The values were
obtained from \url{https://www.finanzen.net/}, accessed on 21 September 2019.} Its underlying asset is the DAX index, and the strike price and maturity are given by  $strike=$ 10,275  and $T=180$ days, respectively.

\begin{figure}[H]
\centering
\begin{subfigure}{8cm}
	\centering
	\includegraphics[width = 8cm]{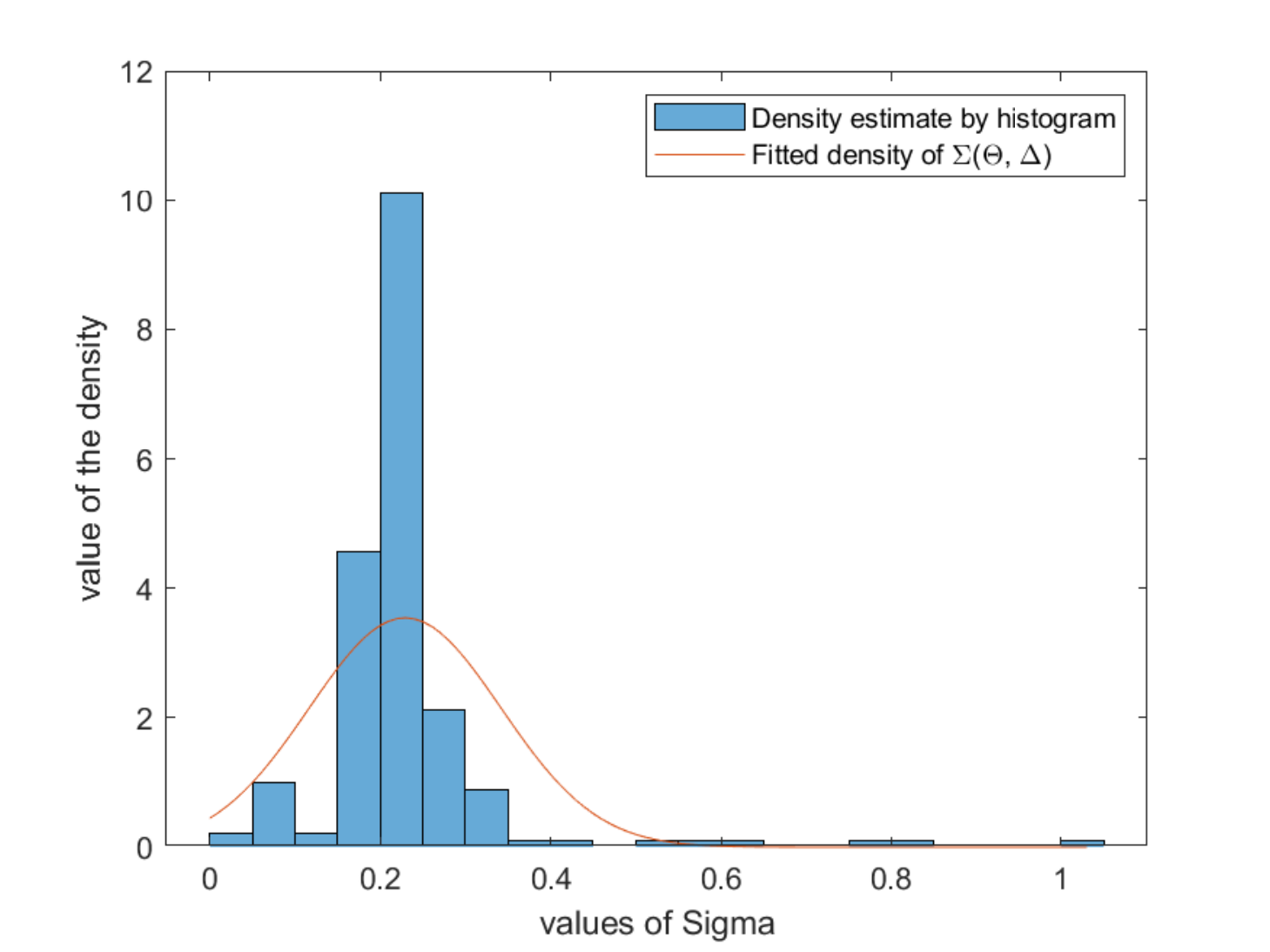}
	\caption{}\label{fig:HermLegfitSigm}
\end{subfigure}~
\begin{subfigure}{8cm}
	\centering
	\includegraphics[width = 8cm]{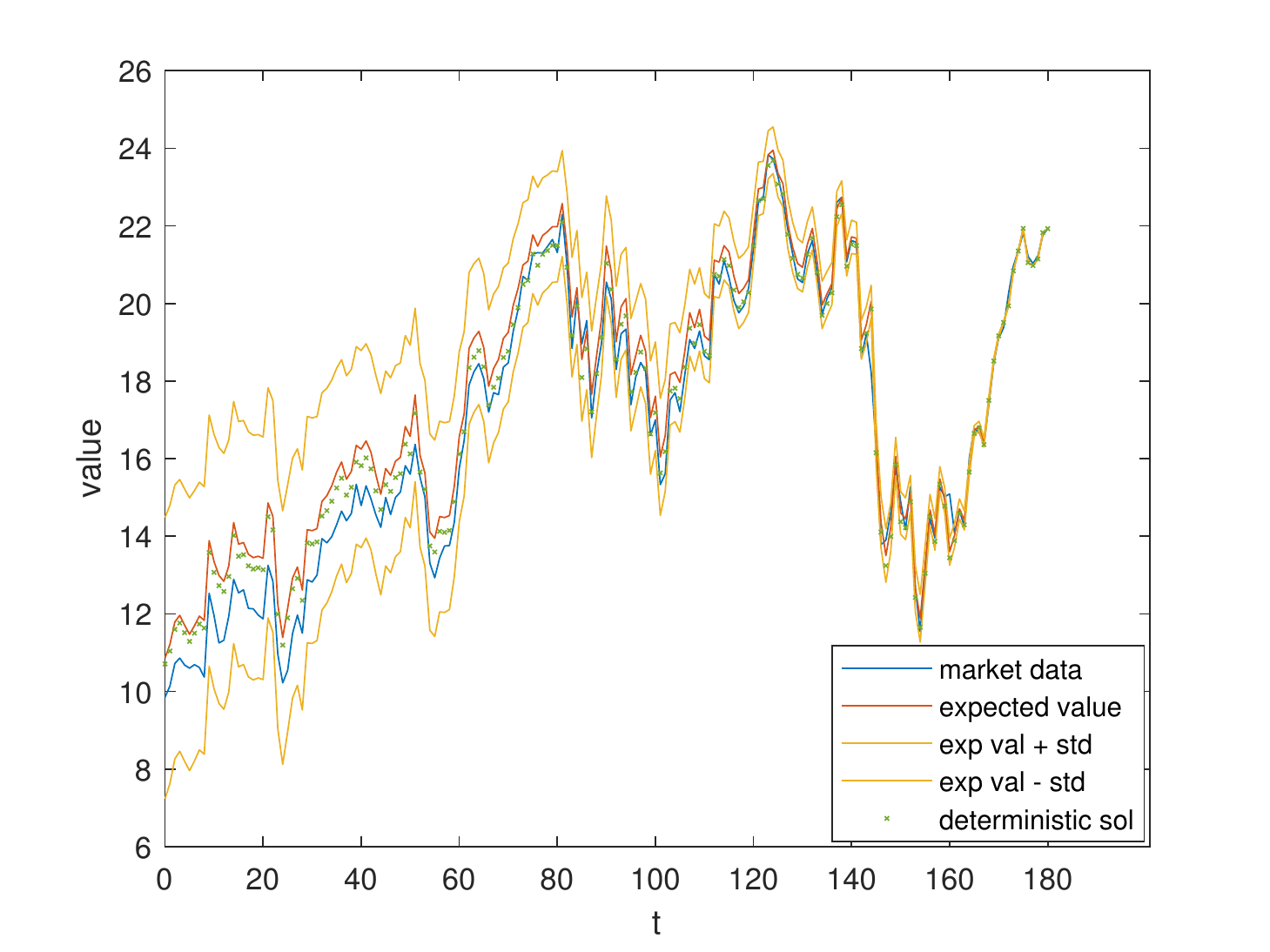}
	\caption{}\label{fig:HermLegModel}
\end{subfigure}\\
\begin{subfigure}{8cm}
	\centering
	\includegraphics[width = 8cm]{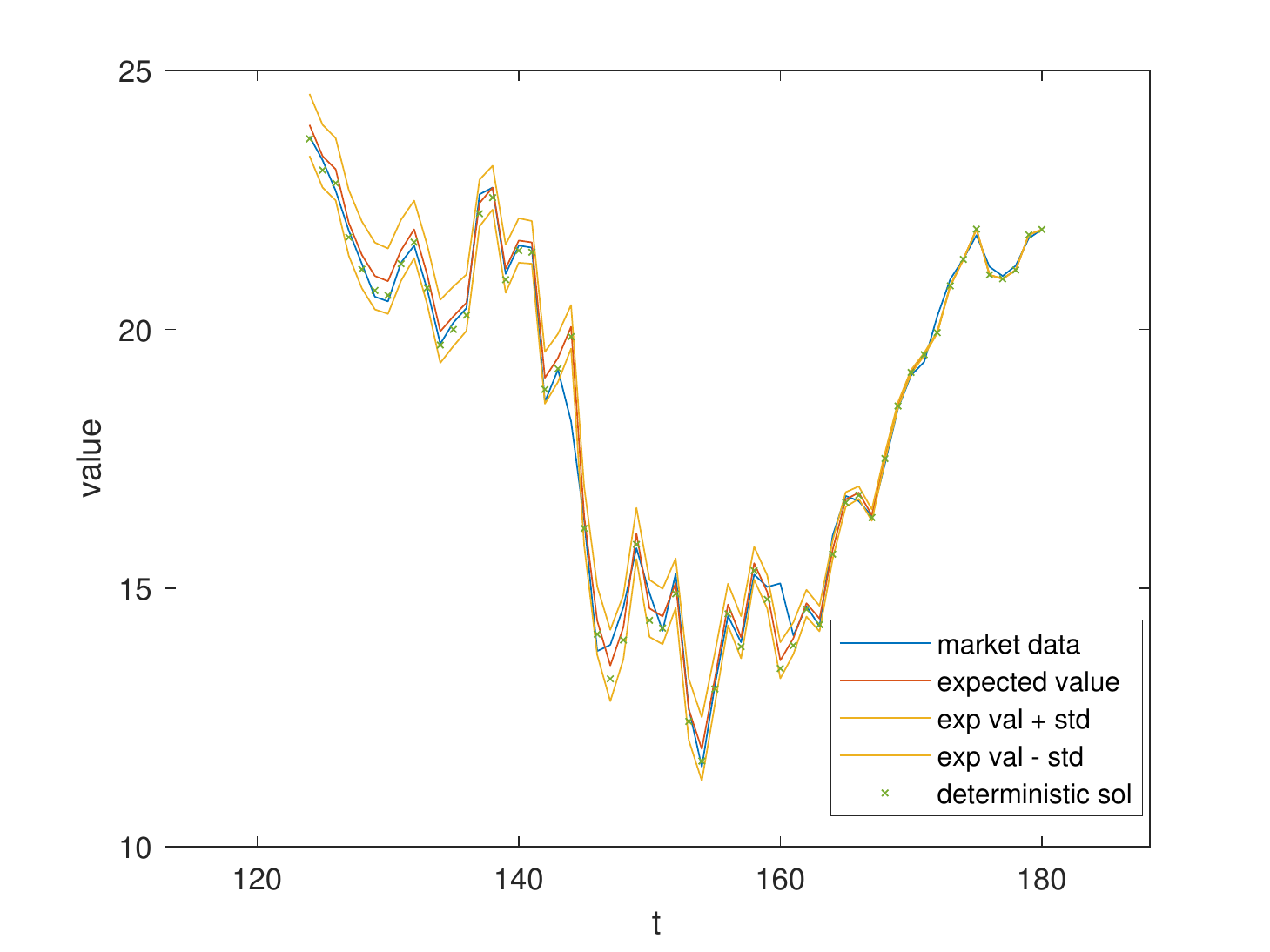}
	\caption{}\label{fig:HermLegModelCloser}
\end{subfigure}
\caption{Comparing the stochastic model to real market data. (\textbf{a}) Histogram
density estimator and density of $\Sigma(\Theta, \Delta)$ fitted to the implied volatilities by constrained maximum likelihood. (\textbf{b}) Market values of the option together with the expected value of the SG solution and the range expected value plus minus standard deviation. (\textbf{c}) Detailed look on the last 55 days.}
\label{fig:MarketComparison}
\end{figure}

A volatility model of the form  $\Sigma(\Theta, \Delta) = \sigma_{00}+\sigma_{10}\Theta+\sigma_{01}\Delta$ was fitted to the daily implied volatilities by using the moment constrained maximum likelihood approach from Example \ref{ex:getSigma} with the two moments mean and variance.
This lead to the volatility model 
\begin{equation}\label{eq:DataModelMultivar}
\Sigma(\Theta, \Delta)=0.2292 + 0.1126\Theta +0.0115 \Delta,
\end{equation}
whose fitted density is shown in Figure \ref{fig:MarketComparison}a together with a histogram density estimator.
The SG solution was computed using the truncation number $N=5$ and the numbers of grid points $M_{\zeta}=200$ and $N_{\tau}=678$. With these values, the numerical scheme is stable and  the system of Equation \eqref{eq:truncateSGsystemMultivar} is parabolic.

Figure \ref{fig:MarketComparison}b shows the market prices and the expected value of the SG solution as well as the range expected value plus/minus standard deviation and the solution to the deterministic Black Scholes equation with volatility $\sigma = E(\Sigma(\Theta, \Delta))$. A more detailed plot of those graphs for the last 55 days of the option is given in Figure \ref{fig:MarketComparison}c. One observes that the expected value of the SG solution is very close to the data on these days but slightly above the data at earlier times. However, the data are always in the range expected value plus/minus standard deviation, as one would expect from stochastic theory.

A comparison to the deterministic solution  shows that it also lies above the market data for early times. Recall that unlike the deterministic solution, the SG solution allows realizations to differ from the expected value within a certain range. 

\subsection{Comparing Bi-Fidelity Solution and High Fidelity Solution}
The Bi-Fidelity solution of the Black Scholes equation with uncertain volatility \eqref{eq:BSestochVolLRV} following volatility model \eqref{eq:volModel} for a European Call option is compared to its high fidelity solution. After that, a simulation is performed to find  the error in expected value and in variance between the Bi-Fidelity solution and the high fidelity solution. The error is characterized in size and shape by a Monte Carlo simulation for different volatility models.  Finally, the computation times for high fidelity and Bi-Fidelity model are compared.

We go back to the toy model of a market with interest rate $r=0$ and  maturity $T=23$ of the option.  The strike price was set to $strike=100$ and the gPC expansion of the solution was truncated after a total polynomial degree of $N=5$ as before.

A rather coarse grid with $M_{\zeta}^L=50$ and $N^L_{\tau} = 150$ was chosen for the low fidelity model. This $N^L_{\tau}$ is high enough such that the vast majority of all low fidelity computations performed in the examples explained below was stable. In the case of instability, the corresponding sample point was removed from the set of low fidelity sample points. The high fidelity solution was computed on a fine grid with  $M_{\zeta}^H+1=350 +1$ grid points in $S$ direction. The number of grid points  $N^H_{\tau}+1 = 5853+1$ in the time direction was chosen such that all high-resolution computations for important volatility models were stable. 

The low fidelity sample points represented volatility models $\Sigma_i(\Theta, \Delta) = \sigma^{(i)}_{00}+\sigma_{10}^{(i)}\Theta +\sigma_{01}^{(i)}\Delta$ with 
\begin{eqnarray}\label{eq:BiFidSamplePoints}
\sigma_{00}^{(i)}&\in& \left\{0< 0.05\lambda\leq 0.8 \,|\, \lambda\in\mathbb{N}\setminus \{0\}\right\}, \nonumber\\ 
\sigma_{10}^{(i)} &\in& \left\{0.05\lambda \leq \sqrt{\sigma_{00}/2} \,|\, \lambda\in \mathbb{N}_0\right\} \spand \\
\sigma_{01}^{(i)} &\in& \left\{0.05\lambda \leq \sqrt{12(\sigma_{00}/2-\sigma_{10}^2)} \,|\, \lambda\in \mathbb{N}_0\right\}.\nonumber
\end{eqnarray}

The coefficients were chosen such that $Var(\Sigma(\Theta, \Delta))\leq \sigma_{00}^{(i)}/2$.

\begin{figure}[H]
\centering
\begin{subfigure}{8cm}
	\centering
	\includegraphics[width = 8cm]{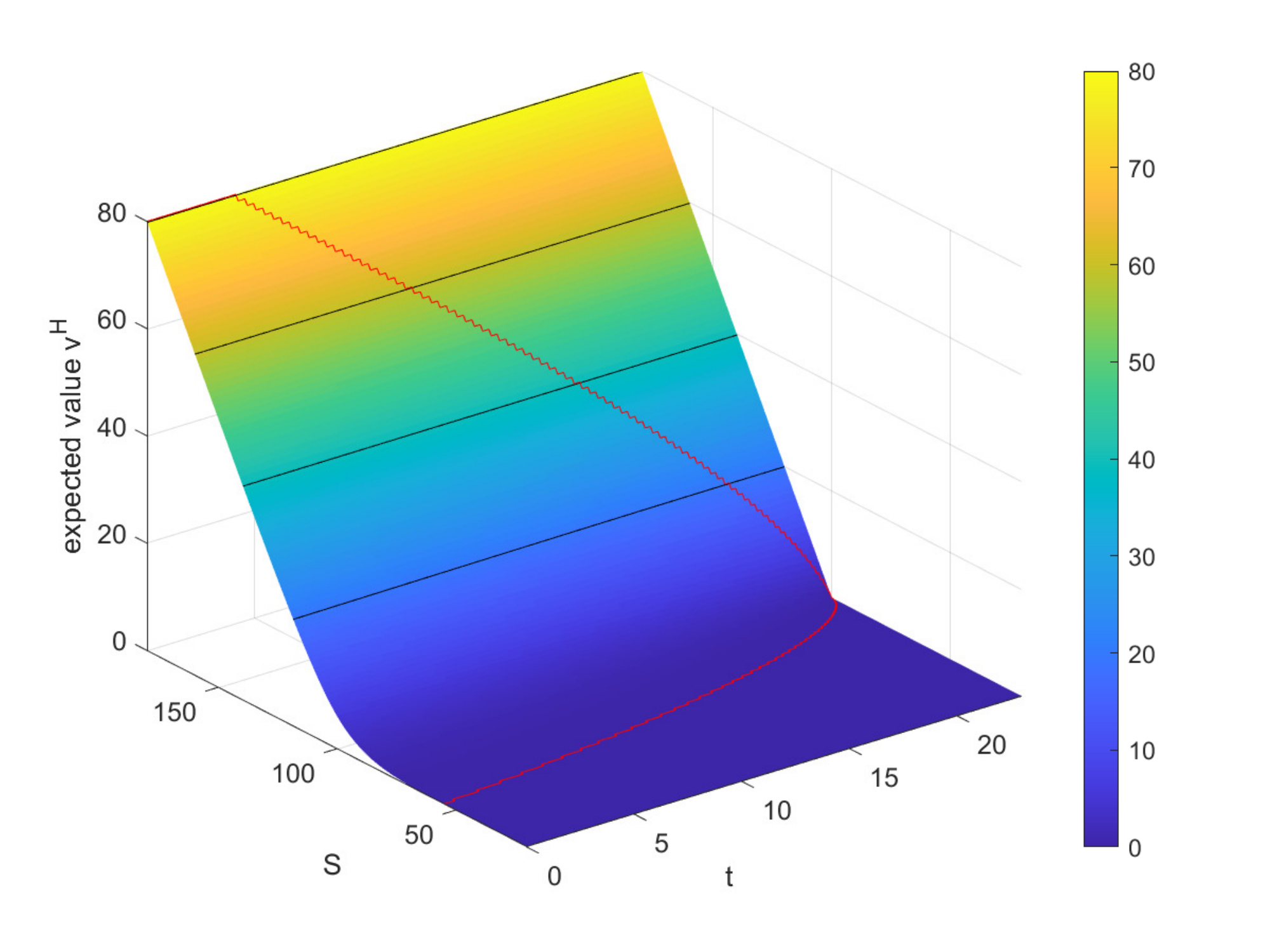}
	\caption{High fidelity solution.}\label{fig:BiFidSig1MeanH}
\end{subfigure}
~
\begin{subfigure}{8cm}
	\centering
	\includegraphics[width = 8cm]{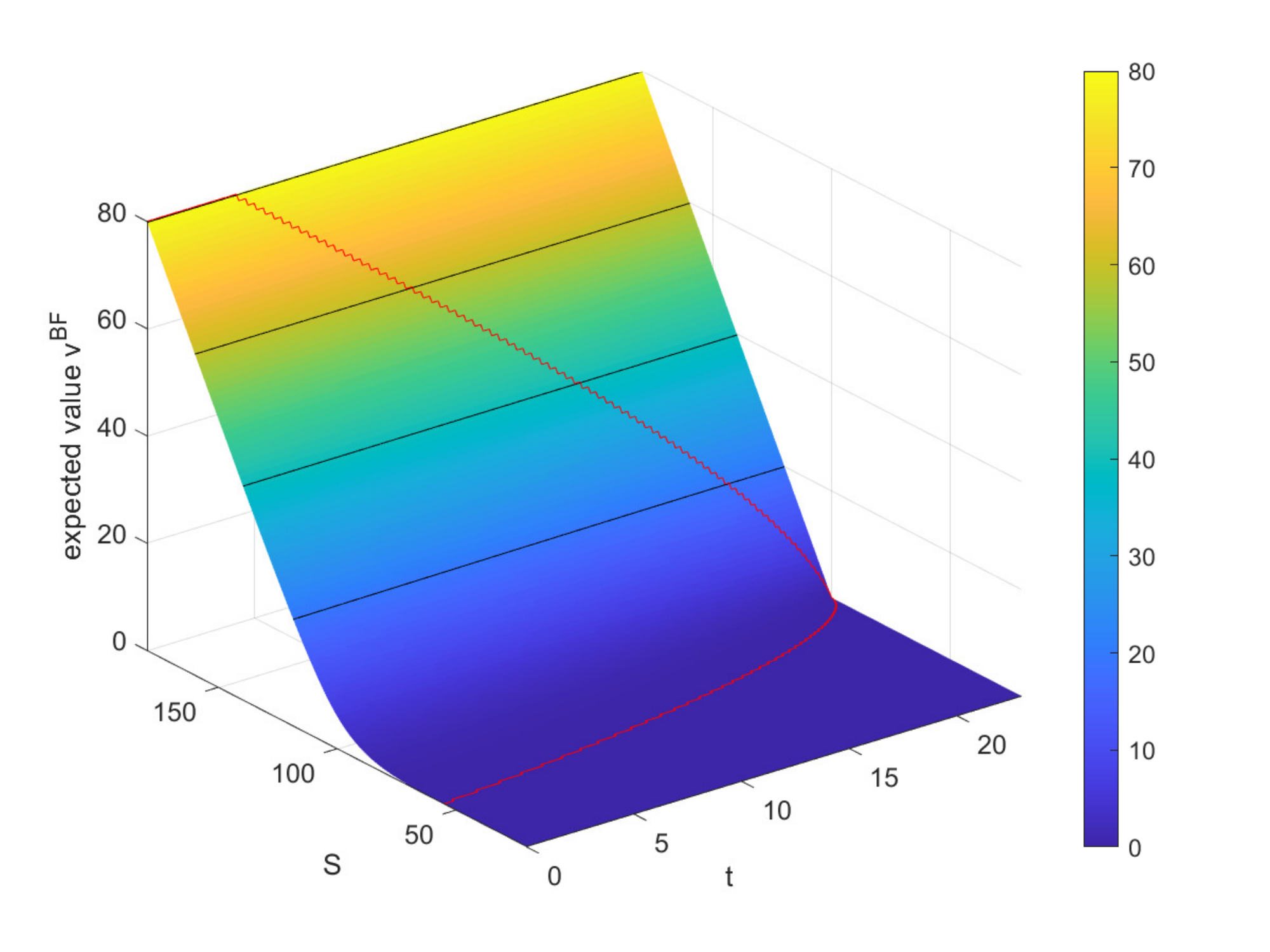}
	\caption{Bi-Fidelity solution.}\label{fig:BiFidSig1MeanBF}
\end{subfigure}
\caption{Expected value surfaces for high fidelity and Bi-Fidelity solution.}
\label{fig:MeanHFBF}
\end{figure}

Figure \ref{fig:MeanHFBF}a,b  shows the expected value surfaces of the high fidelity and the Bi-Fidelity solution for the volatility model $\Sigma(\Theta, \Delta) = 0.5 + 0.2 \Theta + 0.1\sqrt{12}\Delta$. At first glance they seem to approximately coincide.

\begin{figure}[H]
\centering
\begin{subfigure}{8cm}
	\centering
	\includegraphics[width = 8cm]{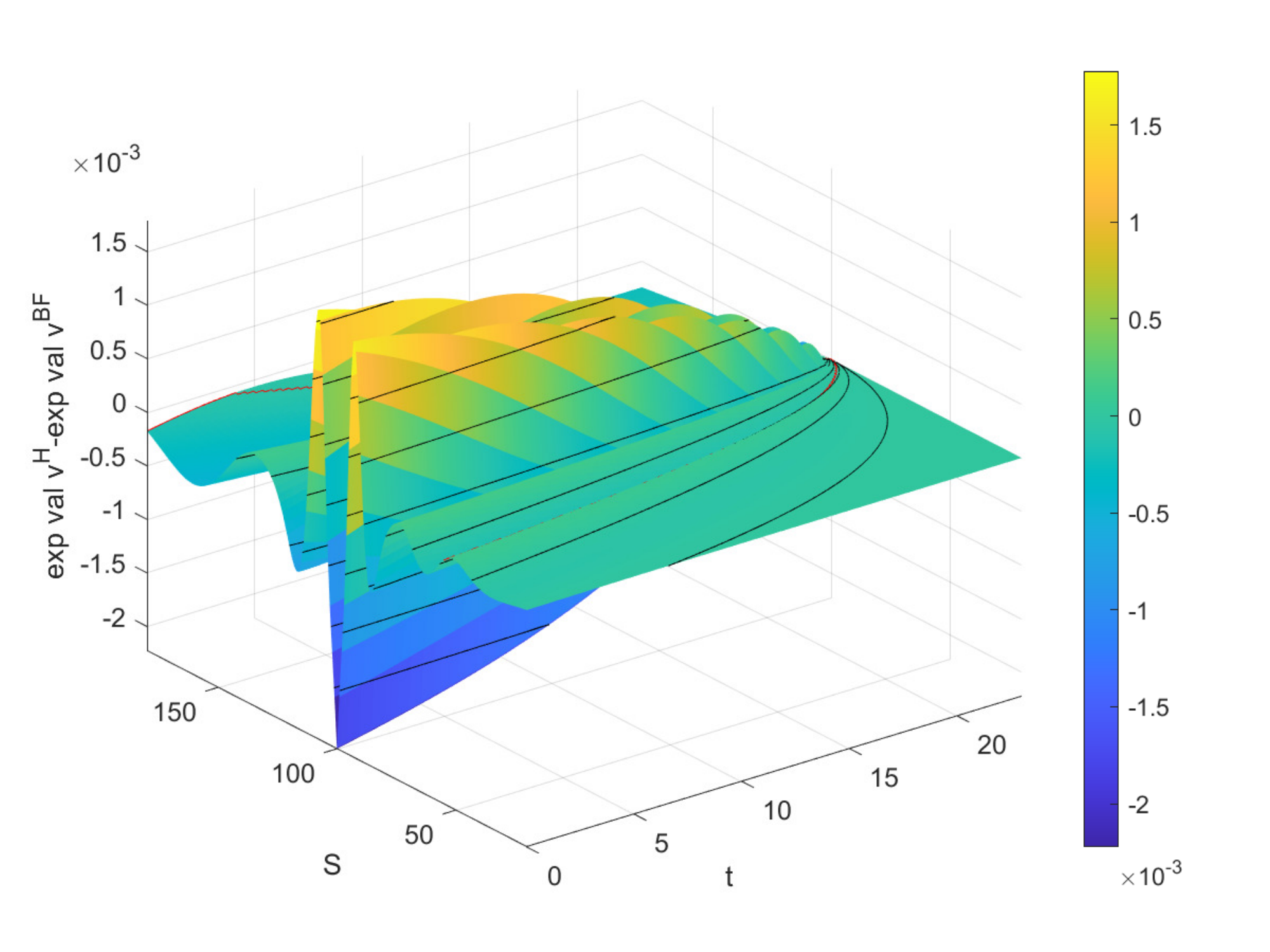}
	\caption{close to the strike price}\label{fig:BiFidSig1Meandiff}
\end{subfigure}
~
\begin{subfigure}{8cm}
	\centering
	\includegraphics[width = 8cm]{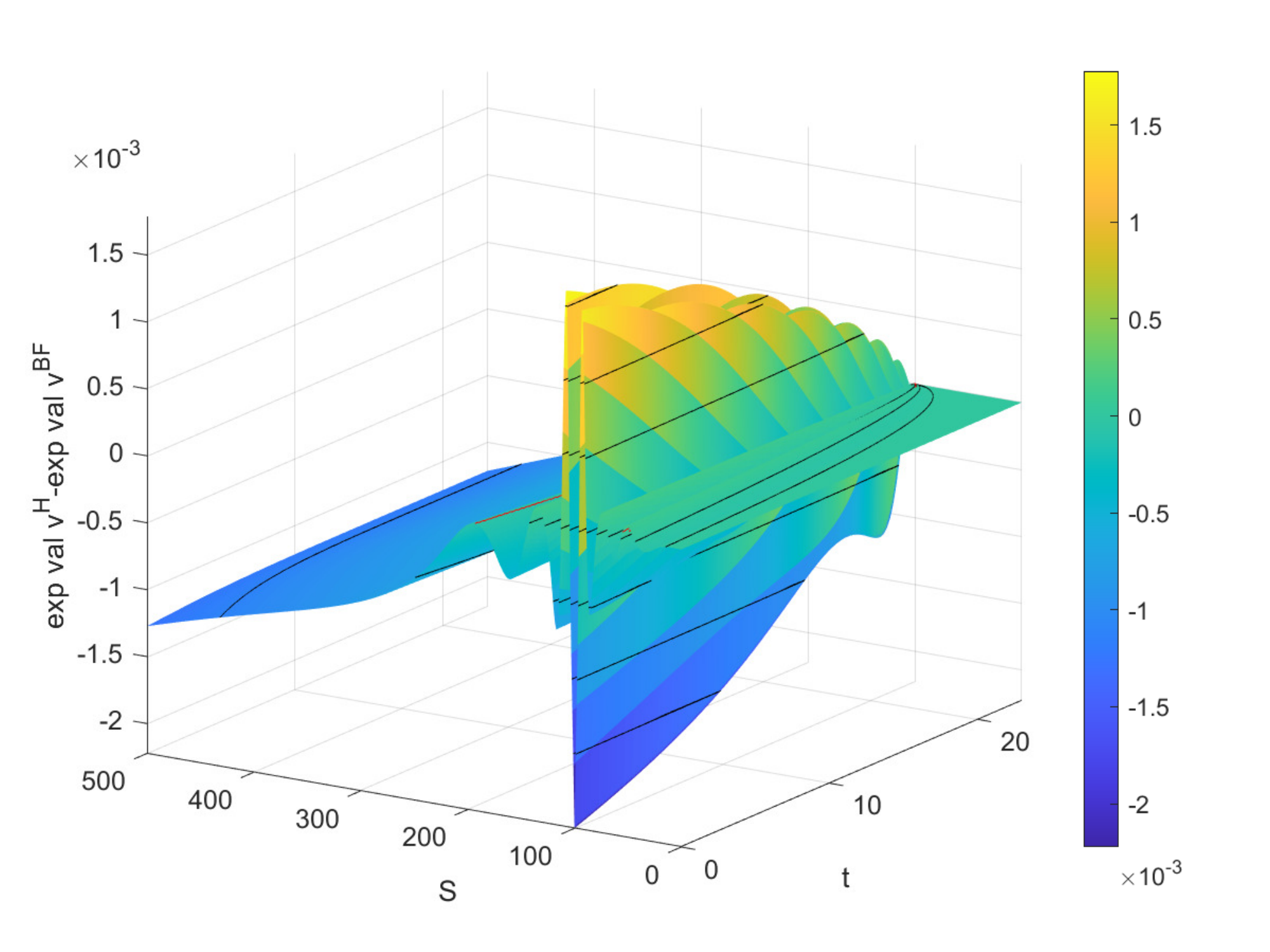}
	\caption{for a wider range of $S$ values}\label{fig:BiFidSig1Meandiff_larger}
\end{subfigure}	\\
\begin{subfigure}{8cm}
	\centering
	\includegraphics[width = 8cm]{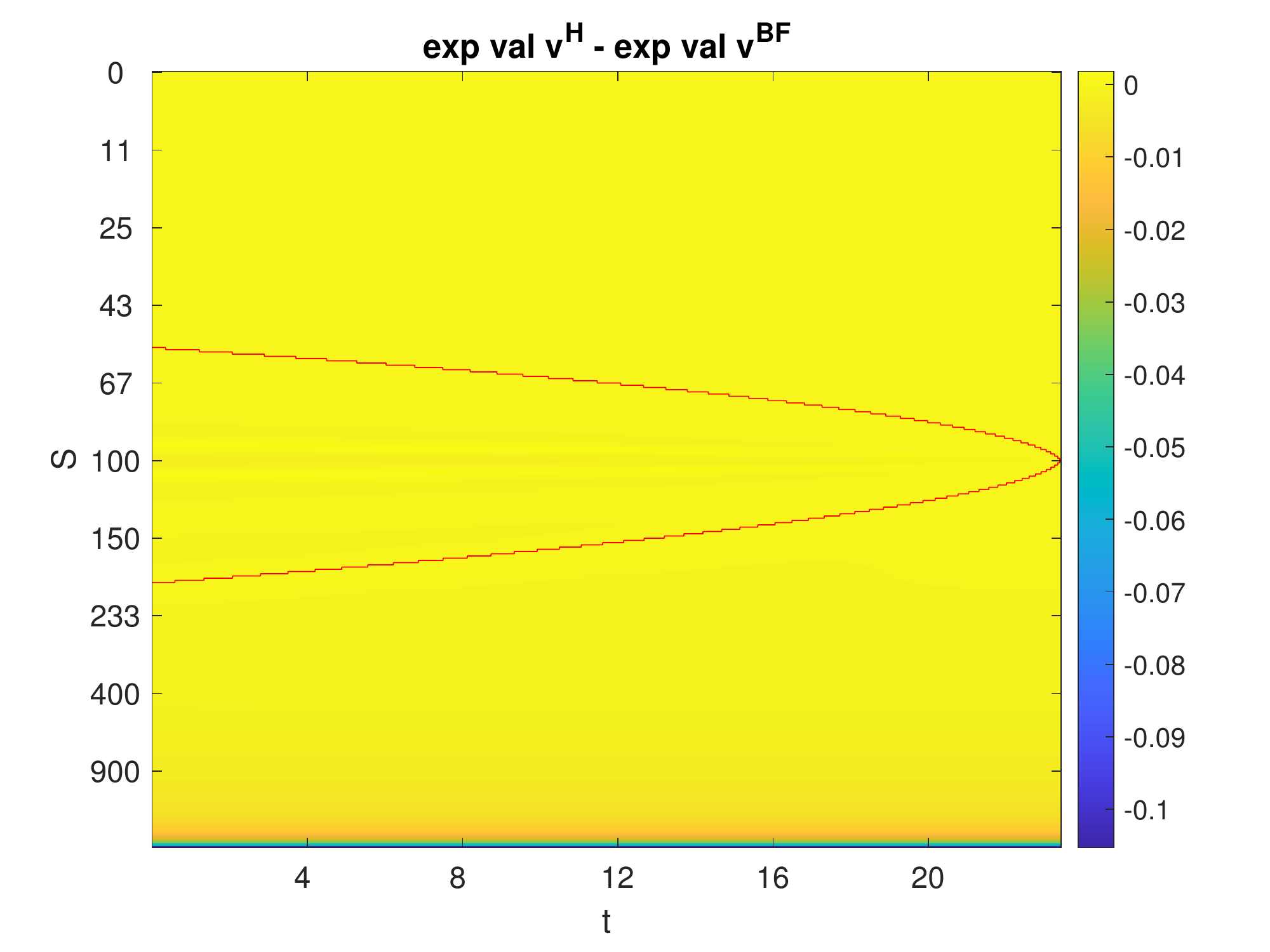}
	\caption{for all $S$ values}\label{fig:BiFidSig1Meandiff_all}
\end{subfigure}
\caption{Absolute
difference in expected value of  high fidelity  and  Bi-Fidelity solution.}
\label{fig:MeanErrorHFBF}
\end{figure}

\begin{figure}[H]
\centering
\begin{subfigure}{8cm}
	\centering
	\includegraphics[width = 8cm]{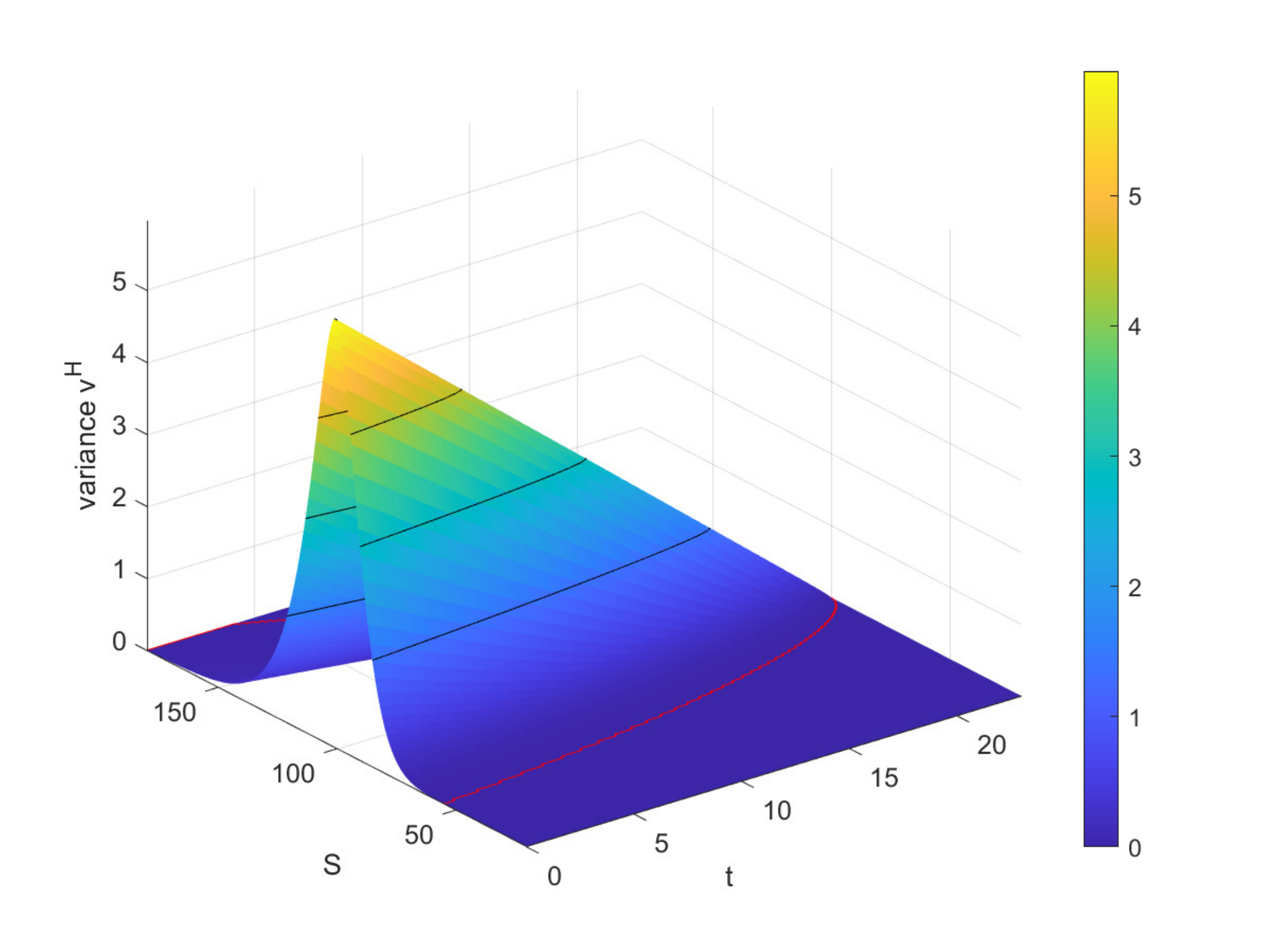}
	\caption{High fidelity solution.}\label{fig:BiFidSig1VarH}
\end{subfigure}
~
\begin{subfigure}{8cm}
	\centering
	\includegraphics[width = 8cm]{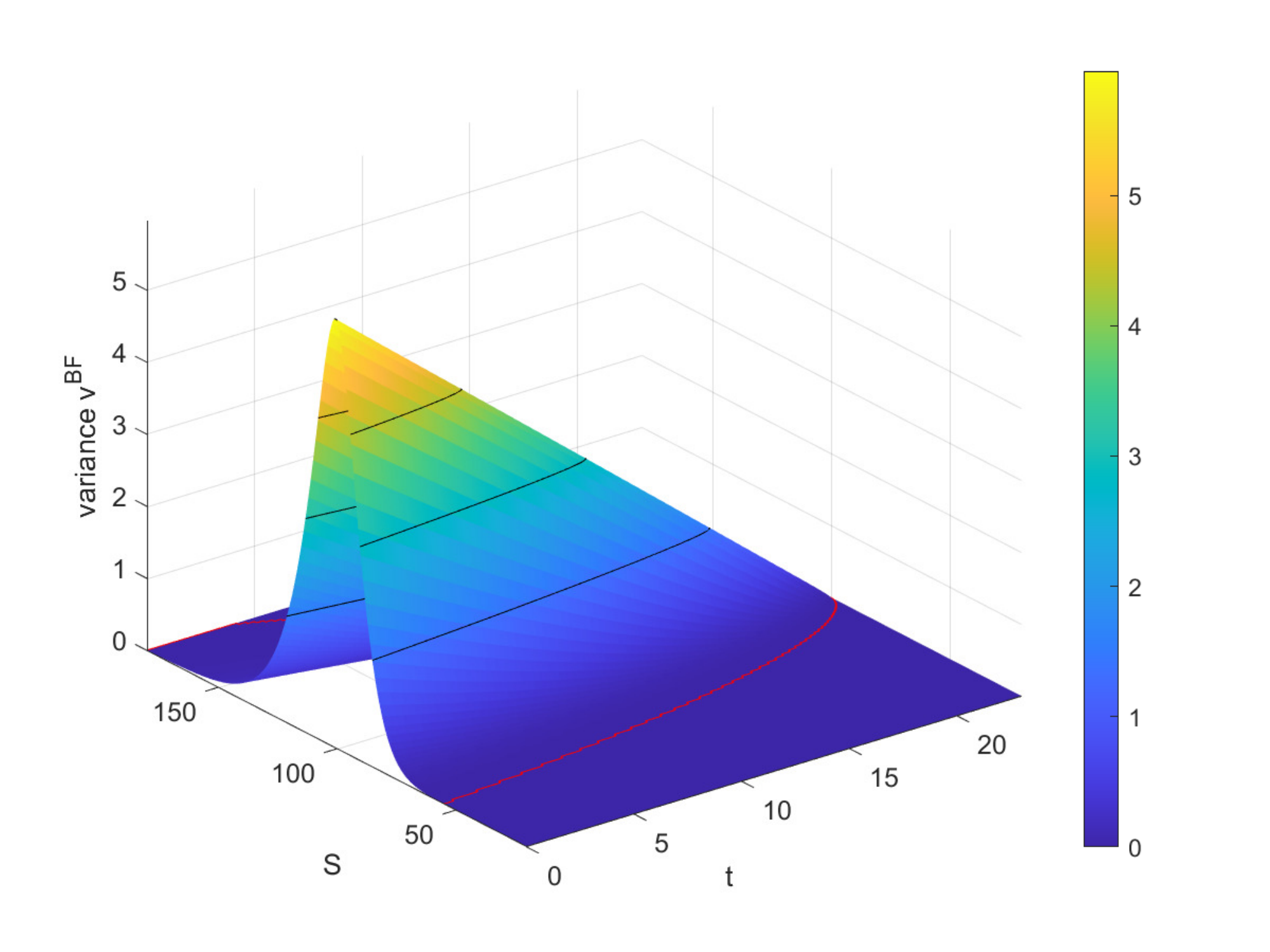}
	\caption{Bi-Fidelity solution.}\label{fig:BiFidSig1VarBF}
\end{subfigure}
\caption{Variance surface for high fidelity and Bi-Fidelity solution.}
\label{fig:VarHFBF}
\end{figure}

To study the deviations, the absolute difference in expected values is displayed in Figure \ref{fig:MeanErrorHFBF}a close to the strike price and Figure \ref{fig:MeanErrorHFBF}b for a wider range of $S$ values. One observes a difference of size $10^{-3}$ within the smoothing area that seems to increase in absolute value as $S\to \infty$.  Figure \ref{fig:MeanErrorHFBF}c shows the  difference for all values of $S$ and $t$. The maximum absolute value of the absolute difference is less than $0.3$ and occurs close to $S=\infty$, where the option values tend to infinity. Therefore, a difference of $0.3$ in these regions means a small deviation. The difference in the smoothing area of size $3\cdot 10^{-3}$
is larger compared to the values attained in this region that are close to zero. 
Recall, however, that the solution is multiplied by $strike$ when transforming back the variables. Hence, an error of size $10^{-3}$ at strike $100$ means an error of size $10^{-5}\cdot strike$. 
\newpage
The variances of high and Bi-Fidelity solution are considered in Figure \ref{fig:VarHFBF}a,b,  respectively. The high fidelity variance seems to be a little bit steeper than the Bi-Fidelity variance.

\begin{figure}[H]
\centering
\begin{subfigure}{8cm}
	\centering
	\includegraphics[width = 8cm]{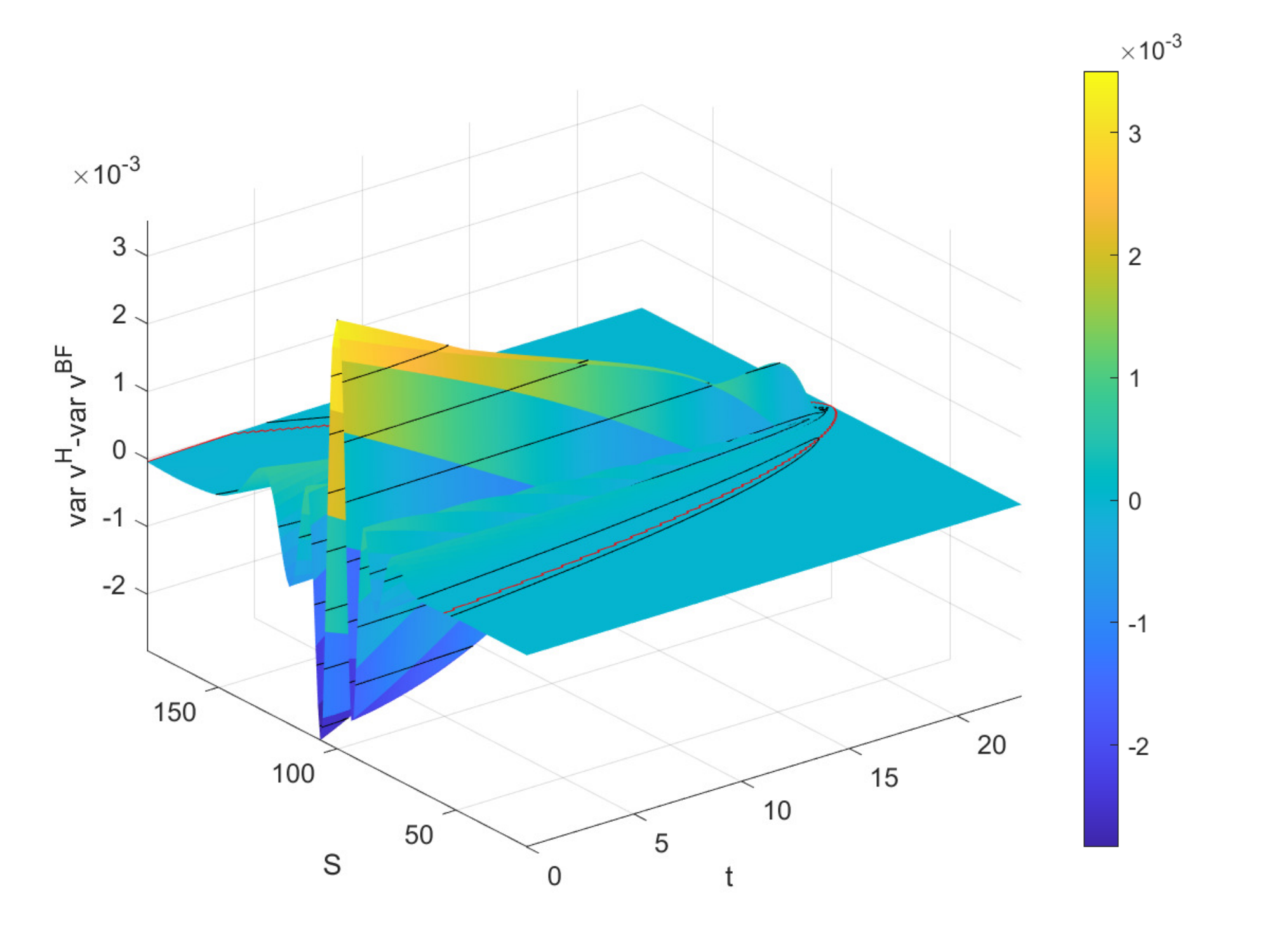}
	\caption{close to the strike pric}\label{fig:BiFidSig1Vardiff}
\end{subfigure}
~
\begin{subfigure}{8cm}
	\centering
	\includegraphics[width = 8cm]{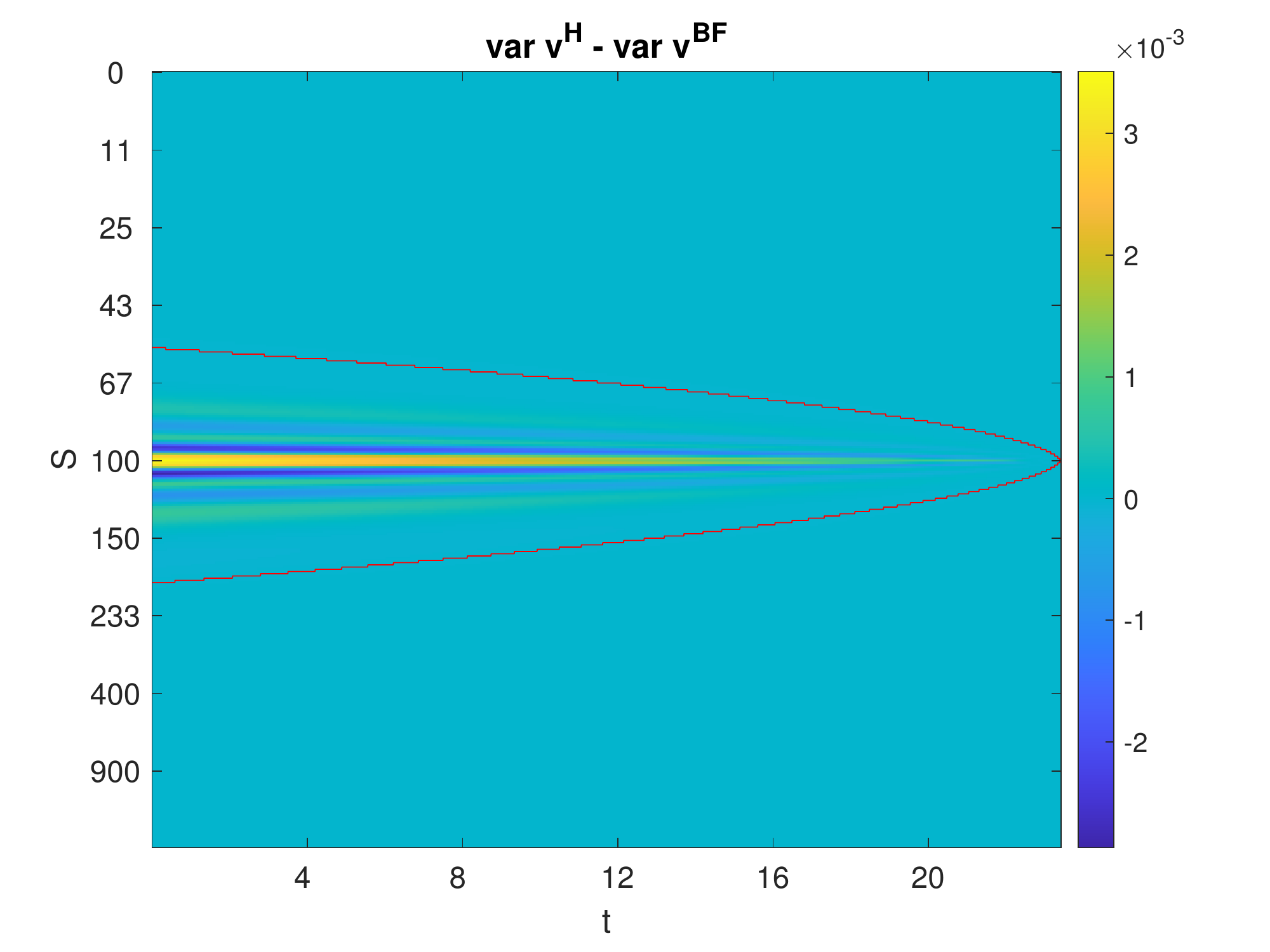}
	\caption{for all $S$ values}\label{fig:BiFidSig1Vardiff_all}
\end{subfigure}
\caption{Absolute
difference in variance  of  high fidelity and  Bi-Fidelity solution.}
\label{fig:VarErrorHFBF}
\end{figure}

We examine the absolute difference in variance as represented in Figure \ref{fig:VarErrorHFBF}a to lie in the smoothing area. Figure \ref{fig:VarErrorHFBF}b, showing the difference for all $S$ and $t$ values, supports this conclusion. The error is again of size $10^{-3}=10^{-7}\cdot strike^2$.

\vspace{-6pt}

Finally, a simulation  was performed to obtain the characteristic size and shape of the Bi-Fidelity error. For this purpose, $300$ volatility models of the form $\Sigma(\Theta, \Delta)=\sigma_{00}+\sigma_{10}\Theta+\sigma_{01}\Delta$ were generated randomly from uniformly distributed random variables $\sigma_{00}\in [0,0.8], \sigma_{10}\in[0,\sqrt{\sigma_{00}/2}], \sigma_{01}\in[0,\sqrt{12(\sigma_{00}/2-\sigma_{10}^2}]$ 'covered' by the grid in \eqref{eq:BiFidSamplePoints}. Both high and Bi-Fidelity solutions were calculated for each of these volatility models.

\begin{figure}[H]
\centering
\begin{subfigure}{8cm}
	\centering
	\includegraphics[width = 8cm]{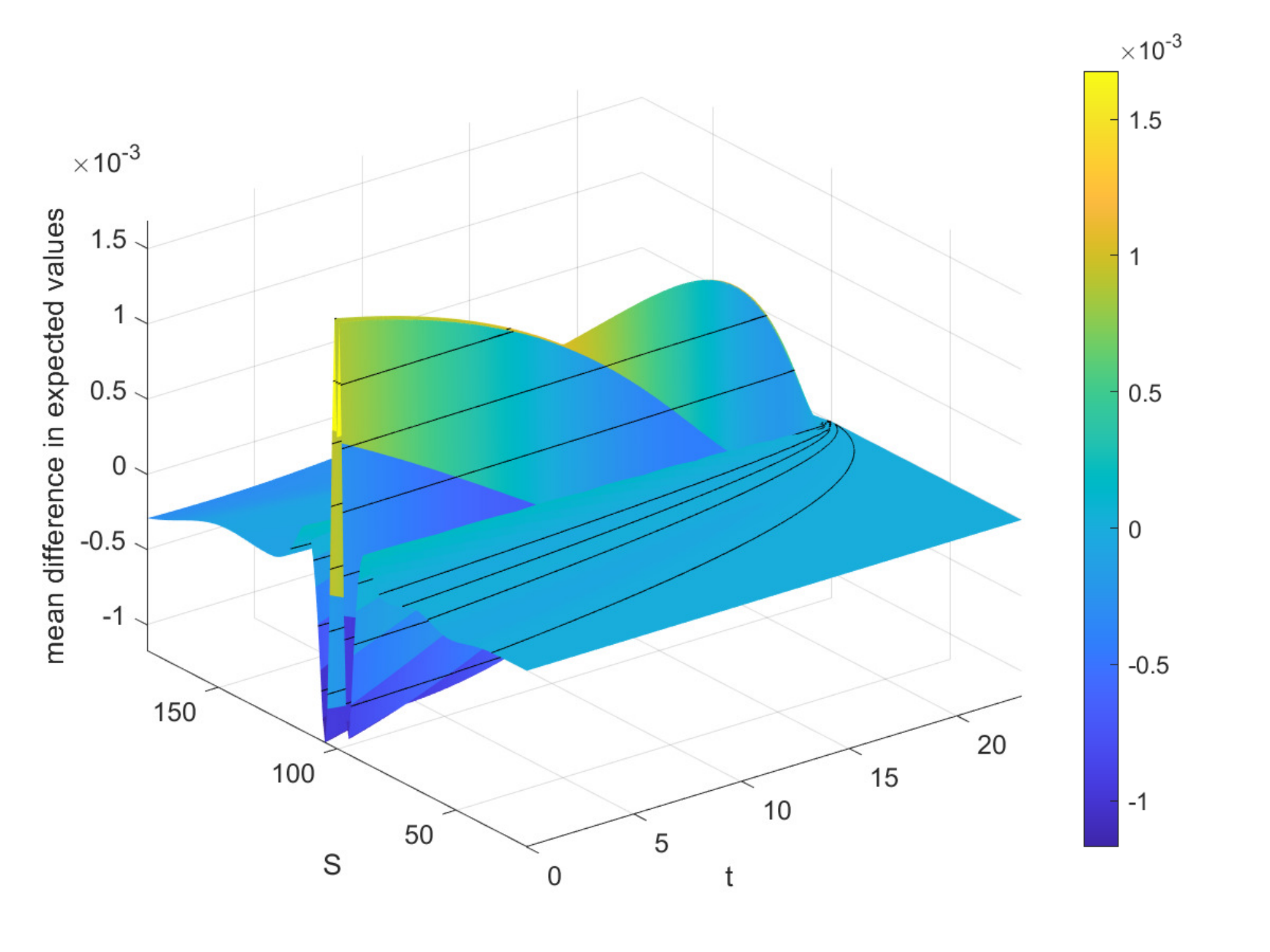}
	\caption{close to the strike price}\label{fig:BiFidMeanMeandiff}
\end{subfigure}
\begin{subfigure}{8cm}
	\centering
	\includegraphics[width = 8cm]{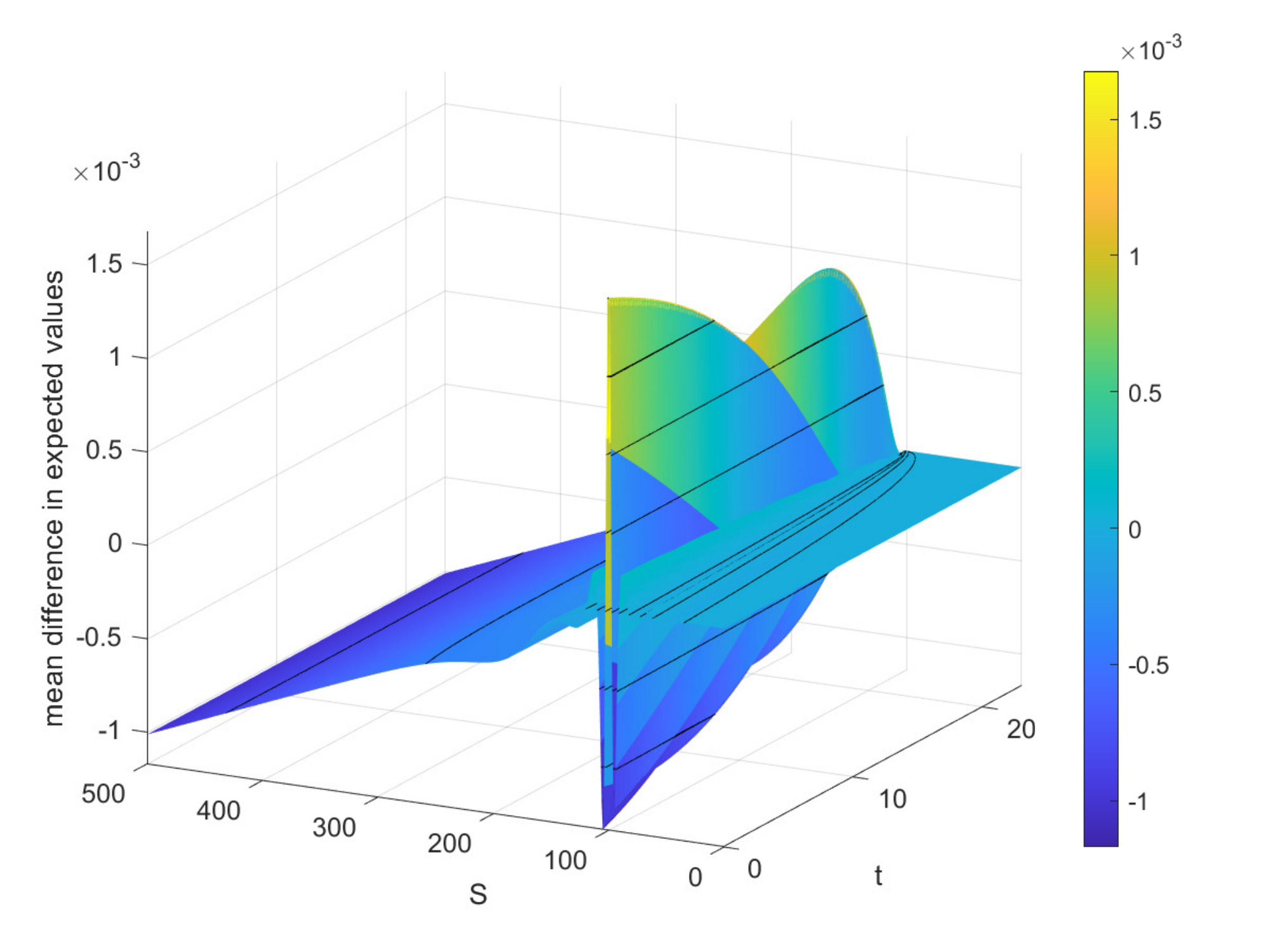}
	\caption{for a wider range of $S$ values}\label{fig:BiFidMeanMeandiff_larger}
\end{subfigure}
\begin{subfigure}{7.5cm}
	\centering
	\includegraphics[width = 7.5cm]{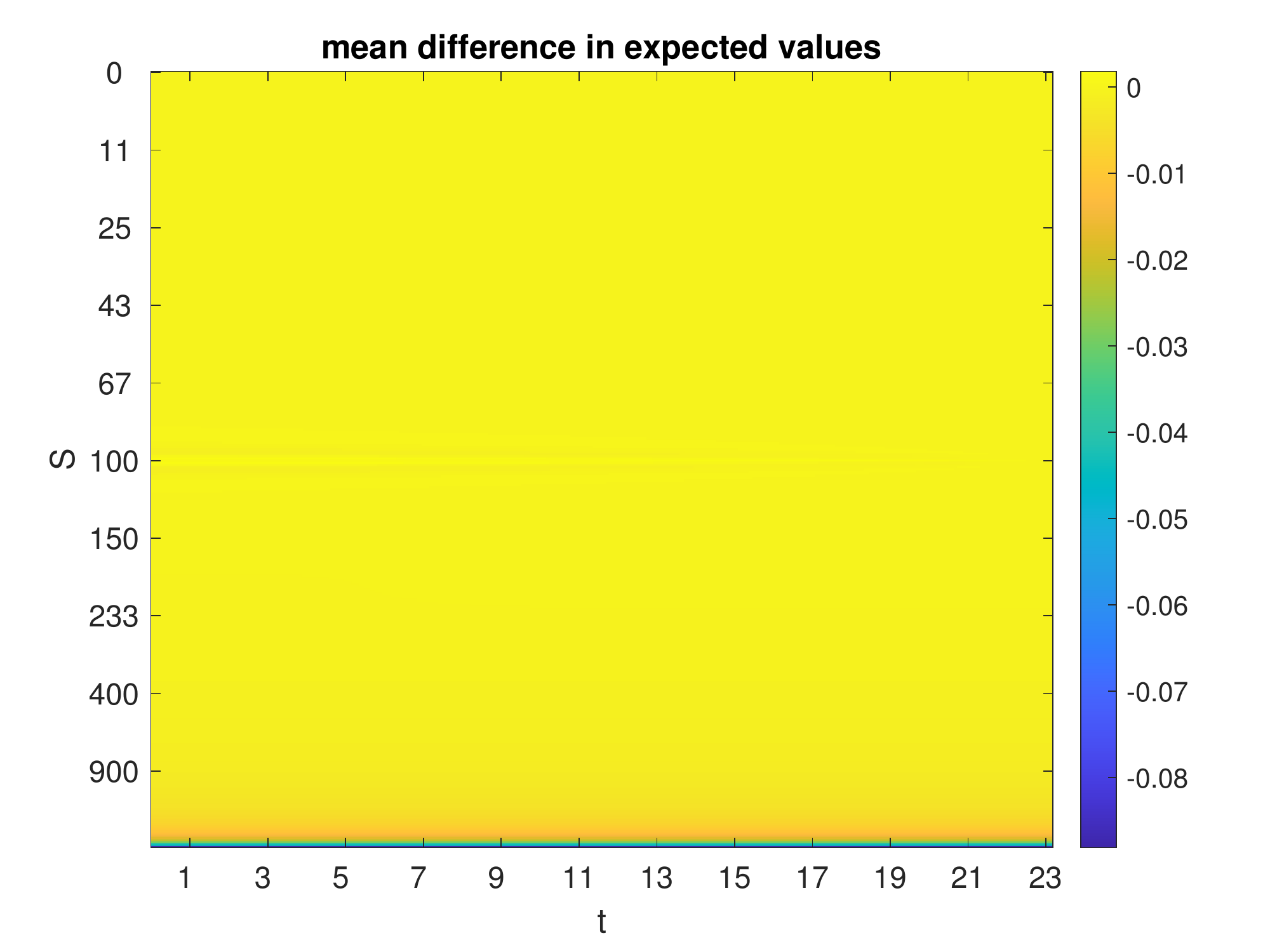}
	\caption{for all $S$ values}\label{fig:BiFidMeanMeandiff_all}
\end{subfigure}
\caption{Mean absolute
difference in expected value of  high fidelity and  Bi-Fidelity solution.}
\label{fig:MCMeanError}
\end{figure}

The mean absolute difference of the expected values is represented in Figure \ref{fig:MCMeanError}a close to the strike price and Figure \ref{fig:MCMeanError}b for a larger range of $S$ values. Figure \ref{fig:MCMeanError}c is a plot of the error for all $S$ and $t$ values. The smoothing area is not plotted since it differs for every volatility model. The shape of the error is characterized by an oscillation of size $10^{-3}=10^{-5}\cdot strike$ close to the strike price and a steady increase in absolute value for $S\to \infty$. The maximum absolute difference lies close to $S=\infty$ and has a size of $10^{-2}= 10^{-4}\cdot strike$, which is small in relative terms. This coincides with the error shape in Figure \ref{fig:MeanErrorHFBF}a--c  and thus seems to be characteristic for the considered Bi-Fidelity model.

\begin{figure}[H]
\centering
\begin{subfigure}{8cm}
	\centering
	\includegraphics[width = 8cm]{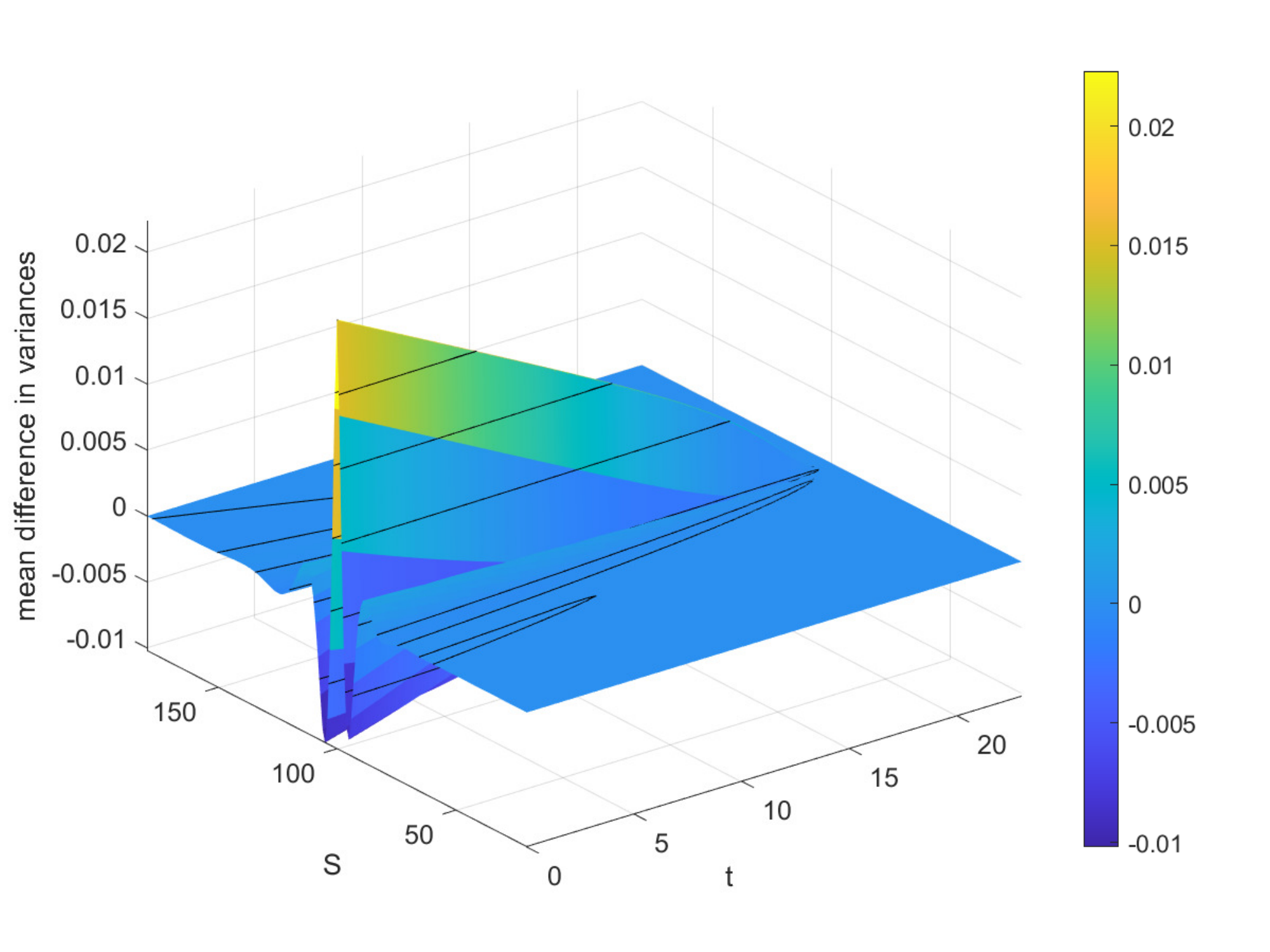}
	\caption{within the smoothing area}\label{fig:BiFidVarMeandiff}
\end{subfigure}~
\begin{subfigure}{8cm}
	\centering
	\includegraphics[width = 8cm]{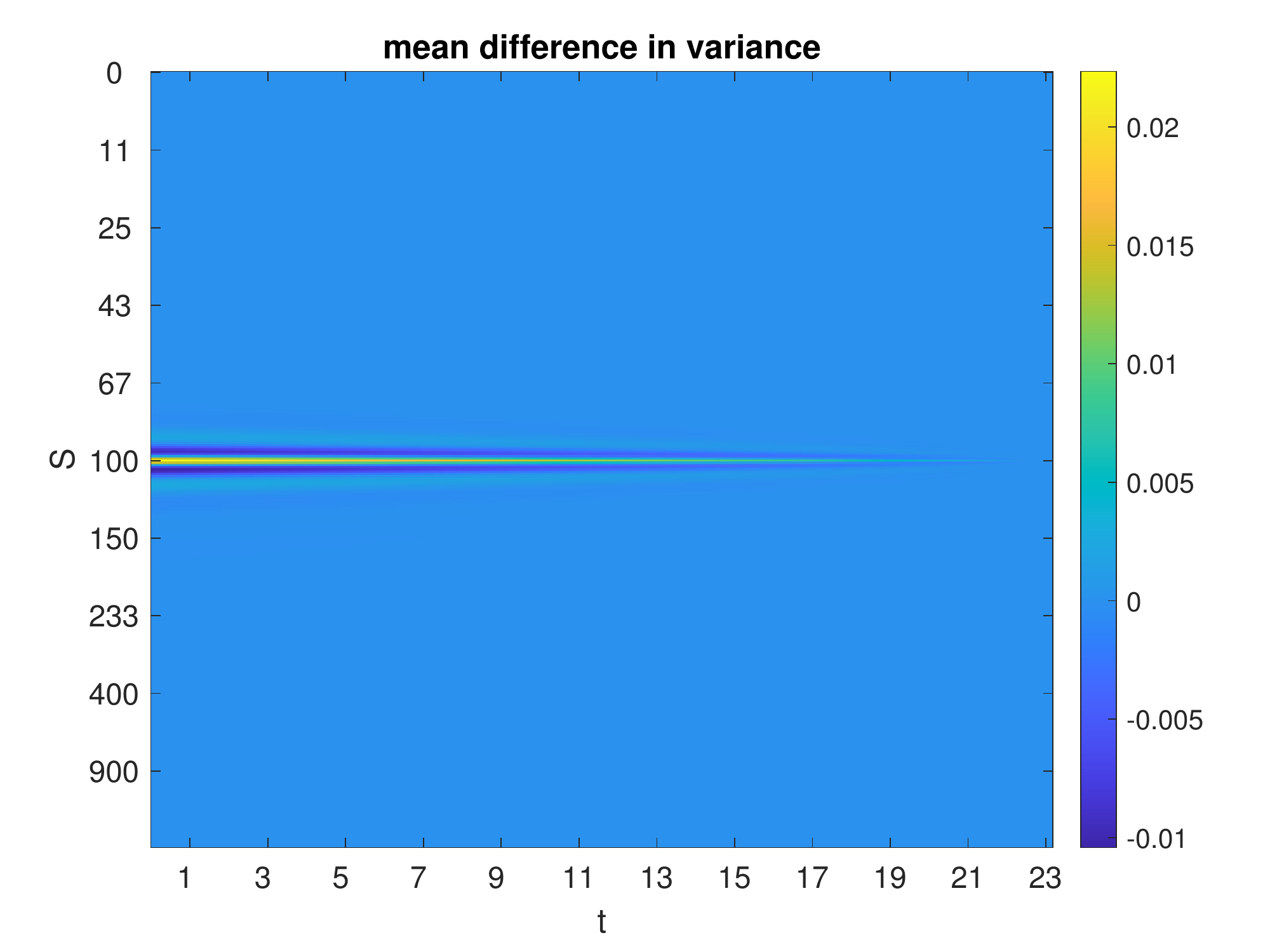}
	\caption{for all $S$ values}\label{fig:BiFidVarMeandiff_all}
\end{subfigure}
\caption{Mean absolute
difference in variance of  high fidelity and  Bi-Fidelity solution.}
\label{fig:MCVarError}
\end{figure}

The characteristic error in the variances derived by the same $300$ volatility models is displayed in Figure \ref{fig:MCVarError}a. It shows some oscillation close to the strike price of size \linebreak\mbox{$10^{-2}= 10^{-6}\cdot strike^2$} but vanishes elsewhere, as one can observe in Figure \ref{fig:MCVarError}b.
This error can possibly be reduced by adding more approximation pairs for the Bi-Fidelity computation  or by  choosing  a low fidelity model closer to the high fidelity model.

\begin{remark}
	This paper focusses on the illustration of the general technique. Hence, the above error analysis is only done for a European Call option as an example. 
	The choice of hyper parameters like the mesh size of the high and low fidelity models depend a lot on the available computational resources and the required accuracy. Therefore, a general recommendation on the choice of hyper-parameters cannot be given. A general rule, however, predicts that the Bi-Fidelity error should decreases when the difference in the low and high fidelity model diminishes.  However, choosing the low fidelity model very close to the high fidelity model cancels the computational advantage of the technique.  A trade-off has to be made considering the specific situation.
	Also an increasing the number $A$ of approximation data pairs should increase the accuracy up to some point. If $A$ is not prescribed by computational resources, one could, e.g., add new approximation points in \eqref{eq:BiFidPointSelection} until the considered distance is below some threshold for all remaining realizations.
\end{remark}

\subsection{Comparison of Computation Times} 

\begin{figure}[H]
\centering
\includegraphics[width = 8cm]{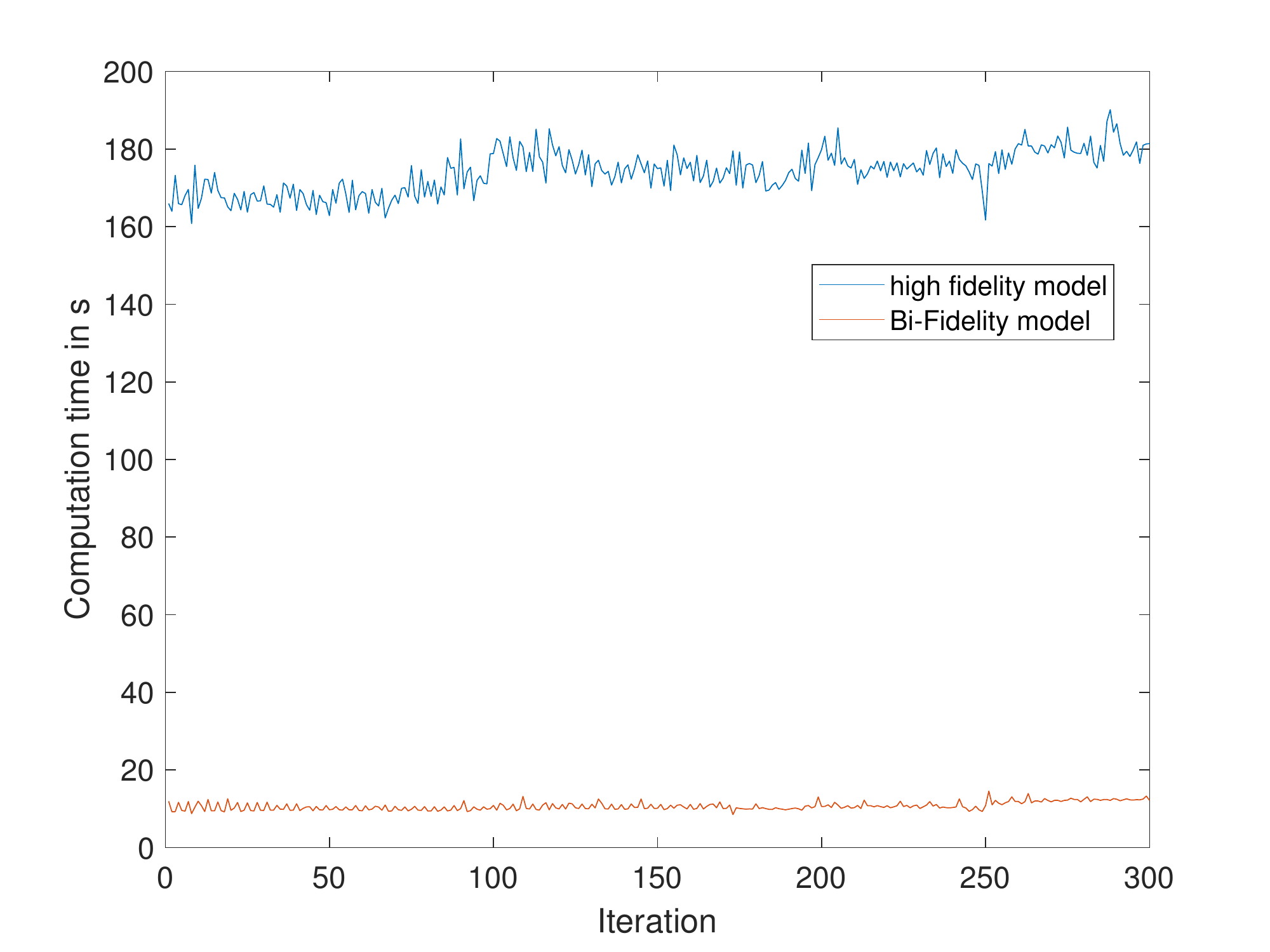}
\caption{Computation times for the high fidelity model and the Bi-Fidelity model evaluated in the same volatility model.}
\label{fig:Computationtime}
\end{figure}

For demonstration, the above Bi-Fidelity model and the high fidelity model with the same number of grid points $M_{\zeta}^H= 350$ and $N_{\tau}^H=5853$ were calculated in the same 300 randomly generated volatility models.  Every model $\Sigma^{(i)}(\Theta, \Delta)=\sigma_{00}^{(i)}+\sigma_{10}^{(i)}\Theta + \sigma_{01}^{(i)}\Delta$ belonging to iteration $i\in\{1,\ldots,300\}$ was generated such that it satisfies the same bounds on the coefficients $\sigma_{00}^{(i)}\in(0,0.8], \sigma_{10}^{(i)} \in [0, \sqrt{\sigma_{00}/2}]$ and $\sigma_{01}^{(i)} \in [0, \sqrt{12(\sigma_{00}/2-\sigma_{10}^2)} ]$ as for the low fidelity sample points in \eqref{eq:BiFidSamplePoints}. The $\Sigma^{(i)}$ should thus be 'covered' by the low fidelity sample points, which enables a Bi-Fidelity computation. In every calculation, the stability of the scheme w.r.t. the chosen time step is checked. The computation times for both models are plotted in Figure \ref{fig:Computationtime}.

The mean computation time for the high fidelity model is 173.99 s, whereas the Bi-Fidelity model achieved a mean computation time of 10.68 s per volatility model. Hence, the application of the Bi-Fidelity method accelerated our computations by a factor of $16.3$ in mean. For finer high fidelity  grids, this difference should further increase. However, choosing a finer grid means introducing a larger difference in the high and low fidelity model, which could lead to larger errors.

\section{Conclusions}
When the volatility in the Black Scholes equation is determined by discrete market data,  uncertainty is introduced due to the estimation procedure. We modelled this uncertainty by a dependence on a finite number of random variables representing random factors of influence. A possibility to fit this uncertain volatility to market data was demonstrated. Afterward, the Black Scholes equation with uncertain volatility was used to model the price process of a derivative. Under certain assumptions,  the random volatility and the stochastic solution can be represented by their generalized Polynomial Chaos (gPC) expansions allowing the application of  the stochastic Galerkin method. The resulting deterministic system of PDEs for the gPC coefficients was truncated and solved numerically by a finite difference scheme.

Numerical examples showed that the expected value of this stochastic model fitted real market data in a similar way as a deterministic model. However, the stochastic solution allows deviations from its expected value within a certain range and it can be used for calculations of further stochastic quantities such as the variance of the solution or in risk management applications. 

However, computation can become costly for a large number of random variables or a late truncation. This is due to the fast increase in the number of gPC coefficients. Therefore, a machine learning technique was presented to reduce the computation cost for computing the solutions for different volatility models within the same setting (option type, maturity, interest rate, maximum polynomial degree). The so-called Bi-Fidelity approach approximates a costly solution on the basis of a computationally cheaper solution and some pre-stored costly solutions for wisely selected volatility models.

For a European Call option, the maximum absolute difference in the expected value of the Bi-Fidelity solution to the desired solution was experimentally observed  to be of size $10^{-5}\cdot strike$ in mean close to the strike price and increase to size $10^{-4}\cdot strike$ in mean for $S\to \infty$, where the expected value also tends to $\infty$. The maximum difference in variance attained a value of size $10^{-6}\cdot strike^2$ in mean. Meanwhile, the mean computation time was decreased by a factor of $16.3$.

{A} topic that is still open to further research is the convergence of the truncated gPC expansion of the stochastic solution to the true solution as the truncation number goes to infinity. If convergence is assumed to hold, it is also possible to  solve the deterministic system of PDEs for the gPC coefficients with a different numerical technique and apply the Bi-Fidelity approach to this solution. Furthermore, one could think of applying the technique used in this paper to the Black Scholes equation with uncertain volatility and interest rates, when there are doubts concerning its true value, or to familiar equations such as the Black Scholes equation for multiple assets or the bond equation.  

\section[\appendixname~\thesection]{Finite Difference Scheme}\label{sec:appendix}\label{sec:Num} 
For demonstrative purposes, European Call options with strike price $strike$ and maturity $T$ will be considered to present the finite difference scheme used for solving the system of Equation \eqref{eq:truncateSGsystemMultivar}.  

For an easier implementation, system \eqref{eq:truncateSGsystemMultivar} is rewritten in vector form. 
This is performed via a bijection $\phi$  from the set $\{ 0,\ldots,|I|-1\}$ of positions in the vector to the set of multi indices $I := \{\delta \in \mathbb{N}_0^L \,|\,|\delta|\leq N\}$, as described in Section 5.2 in \cite{XiuUQ}.
Define $\mathbf{v}:= (v^N_{\phi(0)},\ldots,v^N_{\phi(|I|-1)})^T$, then one can represent system \eqref{eq:truncateSGsystemMultivar} by
\begin{eqnarray*}
	\mathbf{0}_{|I|}= \frac{\partial \mathbf{v}(S,t)}{\partial t} + \frac{1}{2}S^2 \mathbf{A} \frac{\partial ^2 \mathbf{v}(S,t)}{\partial S^2}  + rS \frac{\partial  \mathbf{v}(S,t)}{\partial S}  - r \mathbf{v}(S,t),
\end{eqnarray*}
where the coupling matrix $\mathbf{A}$ is given by 
\begin{eqnarray} \label{eq:defAMultivar}
	\mathbf{A} [n,l] = \sum_{\substack{\alpha, \beta \in \mathbb{N}_0^L, \\ |\alpha|,|\beta|\leq K}} \sigma_{\alpha}\sigma_{\beta} M_{\alpha\beta(\phi(l))(\phi(n))},\hspace{0.5cm} \text{for } n,l=0,\ldots,|I|-1.
\end{eqnarray}

The boundary conditions and final values have to be transformed to vectors as well. If the deterministic part with multiindex $(0,\ldots,0)$ is in the first vector  position and a European Call option is considered, they are given by
\begin{alignat*}{3}
	&\mathbf{v}(S,T)= \begin{pmatrix}
		(S-strike)^+\\ 0\\\vdots\\0
	\end{pmatrix},&&\hspace{0.1cm} S\in (0,\infty), \nonumber\\
	&\mathbf{v}(S,t)\xrightarrow{S\to 0} \mathbf{0}_{|I|},&& \hspace{0.1cm} t\in [0,T], \hspace{0.5cm}\text{ and }\\
	&\frac{1}{S}\mathbf{v}(S,t) \xrightarrow{S\to \infty} \begin{pmatrix}
		1\\ 0\\\vdots\\0
	\end{pmatrix}, &&\hspace{0.1cm} t\in [0,T].\nonumber
\end{alignat*} 

The system has to be transformed into a finite domain. For the European Call option, this can be achieved by the following transformation of variables
\begin{align*}
	&\zeta := \frac{S}{S+strike},\\
	&\tau := T-t,\\
	&\mathbf{\bar{v}}(\zeta, \tau) := \frac{\mathbf{v}(S,t)}{S+strike}= \frac{(1-\zeta)\mathbf{v}(strike\cdot\zeta/(1-\zeta),T-\tau)}{strike},
\end{align*}
which can be found, e.g., in Chapter 2.2.5 in \cite{Chern}  for the deterministic Black Scholes equation. This leads to a PDE for $\mathbf{\bar{v}}$ given by:
\begin{eqnarray}\label{eq:vectortranftruncSG}
	\frac{\partial \mathbf{\bar{v}}(\zeta, \tau)}{\partial \tau} = \frac{1}{2}\zeta^2(1-\zeta)^2\mathbf{A} \frac{\partial^2 \mathbf{\bar{v}}(\zeta, \tau)}{\partial \zeta^2}+ r\zeta(1-\zeta)\frac{\partial \mathbf{\bar{v}}(\zeta, \tau)}{\partial \zeta} -r(1-\zeta)\mathbf{\bar{v}}(\zeta, \tau),\\
	\zeta\in (0,1), \tau \in [0,T],\nonumber
\end{eqnarray}
with boundary and initial conditions 
\begin{alignat*}{3}
	&\mathbf{\bar{v}}(\zeta,0) = \begin{pmatrix}(2\zeta-1)^+ \\ 0 \\ \vdots \\ 0 \end{pmatrix}, &\hspace{0.5cm}& \zeta\in (0,1),\\
	&\mathbf{\bar{v}}(\zeta,\tau)  \xrightarrow{\zeta\to 0} \mathbf{0}_{|I|}, &&\tau\in [0,T], \hspace{0.5cm}\text{ and}\\
	&\mathbf{\bar{v}}(\zeta, \tau) \xrightarrow{\zeta\to 1}\begin{pmatrix} 1 \\ 0 \\ \vdots \\ 0 \end{pmatrix} ,& &\tau\in [0,T].
\end{alignat*}

In order to solve the system, we choose a finite difference scheme because it is easy to implement for practitioners. An equidistant grid  
\begin{eqnarray*}
	&\zeta_m : = \frac{m}{M_{\zeta}} = m \Delta\zeta,&\hspace{0.3cm} m = 0,\ldots,M_{\zeta}, \\
	&\tau^n:= T\frac{n}{N_{\tau}}= n \Delta\tau, &\hspace{0.3cm}n = 0,\ldots,N_{\tau},
\end{eqnarray*}
with $\Delta\zeta := 1/M_{\zeta}, \Delta\tau := T/N_{\tau}$ was selected. The numbers $M_{\zeta}, N_{\tau} \in \mathbb{N}$ were chosen large enough to represent the solution in a proper way and in the right proportion  to obtain a stable scheme. The partial derivatives are approximated component wise by finite differences, as it was done for the deterministic solution in Chapter 8.1.1 in \cite{Chern}, with
\begin{eqnarray*}
	\text{forward differences for}&\frac{\partial \mathbf{\bar{v}}}{\partial \tau}(\zeta_m, \tau^n) 
	\approx& \frac{\mathbf{\bar{v}}(\zeta_m, \tau^{n+1})-\mathbf{\bar{v}}(\zeta_m, \tau^{n})}{\Delta\tau} \hspace{0.5cm}\text{and}\\
	\text{central differences for}	&\frac{\partial \mathbf{\bar{v}}}{\partial \zeta}(\zeta_m, \tau^n) 
	\approx& \frac{\mathbf{\bar{v}}(\zeta_{m+1},\tau^{n}) -\mathbf{\bar{v}}(\zeta_{m-1}, \tau^{n})}{2\Delta\zeta}\\
	\text{and for}&\frac{\partial^2 \mathbf{\bar{v}}}{\partial \zeta^2}(\zeta_m, \tau^n) 
	\approx& \frac{\mathbf{\bar{v}}(\zeta_{m+1}, \tau^{n})-2\mathbf{\bar{v}}(\zeta_{m}, \tau^{n})+\mathbf{\bar{v}}(\zeta_{m-1}, \tau^{n})}{(\Delta\zeta)^2},
\end{eqnarray*}
for $m = 1,\ldots,M_{\zeta}-1, n = 0,\ldots,N_{\tau}-1$. This yields the explicit finite difference scheme\vspace{-6pt}

	\begin{eqnarray}\label{eq:fdScheme}
		\mathbf{\bar{v}}(\zeta_{m}, \tau^{n+1}) = & \Delta\tau \bigg(&\frac{1}{2} \zeta_m^2 (1-\zeta_m)^2 \mathbf{A} \frac{\mathbf{\bar{v}}(\zeta_{m+1}, \tau^{n})-2\mathbf{\bar{v}}(\zeta_{m}, \tau^{n})+\mathbf{\bar{v}}(\zeta_{m-1}, \tau^{n})}{(\Delta\zeta)^2} \nonumber\\
		&&+ r\zeta_m (1-\zeta_m) \frac{\mathbf{\bar{v}}(\zeta_{m+1}, \tau^{n})-\mathbf{\bar{v}}(\zeta_{m-1}, \tau^{n})}{2\Delta\zeta} - r (1-\zeta_m) \mathbf{\bar{v}}(\zeta_{m}, \tau^{n})\bigg)\\
		&+&\mathbf{\bar{v}}(\zeta_{m}, \tau^{n}), \nonumber 
	\end{eqnarray}

for $m = 1,\ldots,M_{\zeta}-1, n = 0,\ldots,N_{\tau}-1$ with initial value
\begin{alignat*}{3}
	&\mathbf{\bar{v}}(\zeta_m,0) = \begin{pmatrix}(2\zeta_m-1)^+ \\ 0 \\ \vdots \\ 0 \end{pmatrix}, &\hspace{0.5cm}& m = 1,\ldots,M_{\zeta}-1.
\end{alignat*}

The remaining values for $m \in\{ 0,M_{\zeta}\}$, i.e., $\zeta_m \in\{ 0,1\}$, are given by the boundary conditions  $\mathbf{\bar{v}}(0,\tau^n) =\mathbf{0}_{N+1}$ and $\mathbf{\bar{v}}(1,\tau^n) =(1,0,\ldots,0)^T$ for all $n$.

Consistency of the scheme can easily be verified. By the Lax--Richtmyer Equivalence theorem, see for instance \cite{StrikwerdaPDE} Theorem 1.5.1,  convergence of the numerical solution is given, if $M_{\zeta}$ and $N_{\tau}$ are chosen to obtain a stable scheme \eqref{eq:fdScheme} and if the system of Equation \eqref{eq:vectortranftruncSG} is well-posed. Well-posedness is in particular given for a parabolic system, i.e., when all real parts of the eigenvalues of $\mathbf{A}$ are positive. 

The Galerkin multiplication tensor and thus the entries of the coupling matrix $\mathbf{A}$ can be computed by a suitable quadrature method. Gaussian quadrature was used to obtain the  numerical results in Section \ref{sec:Results}.



\vspace{6pt} 



\subsection*{Funding}
K.H. acknowledges
support by the Studienstiftung des deutschen Volkes and the Marianne-Plehn-Programm as well as the Würzburg Mathematics Center for Communication and Interaction (WMCCI).
She  was supported by a scholarship from the Hanns-Seidel-Stiftung and the Max Weber-Programm during her Bachelor's and Master's studies.
%

\bibliographystyle{unsrtnat}

\begin{thebibliography}{999}
	
	\bibitem[Whaley(2007)]{WhaleyHistoryDerivatives}
	Whaley, R.
	\newblock {\em Derivatives: Markets, Valuation, and Risk Management}; Wiley
	Finance; Wiley,
	2007.
	
	\bibitem[Crawford and Sen(1996)]{CrawfordThalesStory}
	Crawford, G.; Sen, B.
	\newblock {\em Derivatives for Decision Makers: Strategic Management Issues};
	Wiley Series in Financial Engineering; Wiley, 1996.
	
	\bibitem[Black and Scholes(1973)]{BSmodel}
	Black, F.; Scholes, M.
	\newblock {The Pricing of Options and Corporate Liabilities}.
	\newblock {\em J. Political Econ.} {\bf 1973}, {\em 81},~638--654.
	
	\bibitem[Merton(1973)]{Mertonmodel}
	Merton, R.C.
	\newblock {The Theory of Rational Option Pricing}.
	\newblock {\em Bell J. Econ. Manag. Sci.} {\bf 1973},
	{\em 4},~141--183.
	
	\bibitem[Cox \em{et~al.}(1985)Cox, Ingersoll, and
	Ross]{CoxRossIngersollInterestRateDerivatives}
	Cox, J.C.; Ingersoll, J.E., Jr.; Ross, S.A.
	\newblock A theory of the term structure of interest rates.
	\newblock {\em Econometrica} {\bf 1985}, {\em 53},~385--407.
	
	\bibitem[Rubinstein(1985)]{RubinsteinVolNotConst}
	Rubinstein, M.
	\newblock {Nonparametric Tests of Alternative Option Pricing Models Using All
		Reported Trades and Quotes on the 30 Most Active CBOE Option Classes from
		August 23, 1976 through August 31, 1978}.
	\newblock {\em  J. Financ.} {\bf 1985}, {\em 40},~455--480.
	
	\bibitem[Scott(1987)]{ScottStochVol}
	Scott, L.O.
	\newblock {Option Pricing when the Variance Changes Randomly: Theory,
		Estimation, and an Application}.
	\newblock {\em { J. Financ. Quant. Anal.}} {\bf
		1987}, {\em 22},~419--438.
	
	\bibitem[G{\"u}nther and J{\"u}ngel(2010)]{Jungel}
	G{\"u}nther, M.; J{\"u}ngel, A.
	\newblock Chapter 4 Die Black-Scholes-Gleichung and 8 Einige weiterf{\"u}hrende Themen. In {\em Finanzderivate mit MATLAB}, {2nd} ed.; Vieweg + Teubner,  2010.
	
	
	\bibitem[Dupire(1994)]{Dupire}
	Dupire, B.
	\newblock Pricing with a smile.
	\newblock {\em Risk } {\bf 1994},{\em 7},~18--20
	.
	
	\bibitem[Coleman \em{et~al.}(1999)Coleman, Li, and Verma]{ColemanEtAlLocalVol}
	Coleman, T.; Li, Y.; Verma, A.
	\newblock Reconstructing the unknown local volatility function.
	\newblock {\em  J. Comput. Financ.} {\bf 1999}, {\em 2},~77--102.
	
	\bibitem[Crepey(2002)]{CrepeyLocalVol}
	Crepey, S.
	\newblock {Calibration of the local volatility in a trinomial tree using
		Tikhonov regularization}.
	\newblock {\em Inverse Probl.} {\bf 2002}, {\em 19},~91--127.
	
	\bibitem[Hanke and R{\"o}sler(2005)]{HankeRoslerLocalVol}
	Hanke, M.; R{\"o}sler, E.
	\newblock {Computation of Local Volatilities from Regularized Dupire
		Equations}.
	\newblock {\em {Int. J. Theor. Appl. Financ.}} {\bf
		2005}, {\em 8},~207--221.
	
	\bibitem[Heston(1993)]{HestonModel}
	Heston, S.L.
	\newblock {A Closed-Form Solution for Options with Stochastic Volatility with
		Applications to Bond and Currency Options}.
	\newblock {\em  Rev. Financ. Stud.} {\bf 1993}, {\em 6},~327--343.
	
	\bibitem[Hull and White(1987)]{HullWhiteStochVol}
	Hull, J.C.; White, A.
	\newblock {The Price of Options on Assets with Stochastic Volatilities}.
	\newblock {\em  J. Financ.} {\bf 1987}, {\em 42},~281--300.
	
	\bibitem[Mishura and Ralchenko(2021)]{DiscreteTime_Mishura2021}
	Mishura, Y.; Ralchenko, K.
	\newblock {\em Discrete-Time Approximations and Limit Theorems: In Applications
		to Financial Markets}; De Gruyter,
	2021.
	\newblock
	doi:\href{https://doi.org/10.1515/9783110654240}{10.1515/9783110654240}.
	
	\bibitem[Namihira and Kopriva(2012)]{KoprivaSCollBS}
	Namihira, M.; Kopriva, D.A.
	\newblock Computation of the effects of uncertainty in volatility on option
	pricing and hedging.
	\newblock {\em Int. J. Comput. Math.} {\bf 2012}, {\em
		89},~1281--1302.
	
	\bibitem[Pulch and van Emmerich(2009)]{PulchBSStochGal}
	Pulch, R.; van Emmerich, C.
	\newblock Polynomial chaos for simulating random volatilities.
	\newblock {\em Math. Comput. Simul.} {\bf 2009}, {\em
		80},~245--255.
	
	\bibitem[Drakos(2016)]{DrakosBSStochGal}
	Drakos, S.
	\newblock Uncertain Volatility Derivative Model Based on the Polynomial Chaos.
	\newblock {\em J. Math. Financ.} {\bf 2016}, {\em 6},~55--63.
	
	\bibitem[Zhang \em{et~al.}(2018)Zhang, Ding, and Scheffel]{zhang2018policy}
	Zhang, Y.; Ding, S.; Scheffel, E.
	\newblock Policy impact on volatility dynamics in commodity futures markets:
	Evidence from China.
	\newblock {\em J. Futur. Mark.} {\bf 2018}, {\em 38},~1227--1245.
	
	\bibitem[Bazzana and Collini(2020)]{BAZZANA2020101240}
	Bazzana, F.; Collini, A.
	\newblock {How does HFT activity impact market volatility and the bid-ask
		spread after an exogenous shock? An empirical analysis on S\&P 500 ETF}.
	\newblock {\em  N. Am. J. Econ. Financ.} {\bf 2020},
	{\em 54},~101240.
	\newblock
	doi:\href{https://doi.org/10.1016/j.najef.2020.101240}{10.1016/j.najef.2020.101240}.
	
	\bibitem[Xie \em{et~al.}(2021)Xie, Cui, and Liu]{xie2021does}
	Xie, D.; Cui, Y.; Liu, Y.
	\newblock How does investor sentiment impact stock volatility? New evidence
	from Shanghai A-shares market.
	\newblock {\em China Financ. Rev. Int.} {\bf 2021}. https://doi.org/10.1108/CFRI-01-2021-0007.
	
	\bibitem[Miloș(2021)]{laws10030055}
	Miloș, M.C.
	\newblock Impact of MiFID II on the Market Volatility---Analysis on Some
	Developed and Emerging European Stock Markets.
	\newblock {\em Laws} {\bf 2021}, {\em 10}, 55.
	\newblock
	doi:\href{https://doi.org/10.3390/laws10030055}{10.3390/laws10030055}.
	
	\bibitem[Sullivan(2015)]{SullivanUQ}
	Sullivan, T.J.
	\newblock {{Introduction to Uncertainty Quantification}}. In {\em
		Texts in Applied Mathematics}; Springer,
	2015; Volume~63.
	
	\bibitem[Rahman(2018)]{RahmangPCDependency}
	Rahman, S.
	\newblock A polynomial chaos expansion in dependent random variables.
	\newblock {\em J. Math. Anal. Appl.} {\bf 2018},
	{\em 464},~749--775.
	
	\bibitem[Janson(1997)]{JansonLpIsom}
	Janson, S.
	\newblock {\em Gaussian Hilbert Spaces}; Cambridge University Press: Cambridge, UK,
	1997.
	
	\bibitem[De \em{et~al.}(2020)De, Maute, and Doostan]{de2020bi}
	De, S.; Maute, K.; Doostan, A.
	\newblock Bi-fidelity stochastic gradient descent for structural optimization
	under uncertainty.
	\newblock {\em Comput. Mech.} {\bf 2020}, {\em 66},~745--771.
	
	\bibitem[Fairbanks \em{et~al.}(2020)Fairbanks, Jofre, Geraci, Iaccarino, and
	Doostan]{FAIRBANKS2020108996}
	Fairbanks, H.R.; Jofre, L.; Geraci, G.; Iaccarino, G.; Doostan, A.
	\newblock Bi-fidelity approximation for uncertainty quantification and
	sensitivity analysis of irradiated particle-laden turbulence.
	\newblock {\em J. Comput. Phys.} {\bf 2020}, {\em 402},~108996.
	\newblock
	doi:\href{https://doi.org/10.1016/j.jcp.2019.108996}{10.1016/j.jcp.2019.108996}.
	
	\bibitem[Zhu \em{et~al.}(2014)Zhu, Xueyu and Narayan, Akil and Xiu,
	Dongbin]{ZhuNarayanXiuStochCollBiFid}
	{Zhu, X.; Narayan, A.; Xiu, D}.
	\newblock {Computational Aspects of Stochastic Collocation with Multifidelity
		Models}.
	\newblock {\em {SIAM/ASA J. Uncertain. Quantif.}} {\bf 2014},
	{\em 2},~444--463.
	
	\bibitem[Narayan \em{et~al.}(2014)Narayan, Gittelson and Xiu]{NarayanGittelsonXiuStochCollBiFid}
	{Narayan, A.; Gittelson, C.; Xiu, D}.
	\newblock {A Stochastic Collocation Algorithm with Multifidelity Models}.
	\newblock {\em {SIAM J. Sci. Comput.}} {\bf 2014}, {\em
		36},~495--521.
	
	\bibitem[Liu and Zhu(2020)]{LiuBiFid}
	Liu, L.; Zhu, X.
	\newblock {A bi-fidelity method for the multiscale Boltzmann equation with
		random parameters}.
	\newblock {\em J. Comput. Phys.} {\bf 2020}, {\em 402}, 108914.
	
	\bibitem[Gamba \em{et~al.}(2021)Gamba, Jin, and Liu]{Gamba_2021}
	Gamba, I.M.; Jin, S.; Liu, L.
	\newblock Error estimates of a Bifidelity method for kinetic equations with
	random parameters and multiple scales.
	\newblock {\em Int. J. Uncertain. Quantif.} {\bf
		2021}, {\em 11},~57--75.
	
	\bibitem[Gao \em{et~al.}(2020)Gao, Zhu, and Wang]{GAO2020113047}
	Gao, H.; Zhu, X.; Wang, J.X.
	\newblock A bi-fidelity surrogate modeling approach for uncertainty propagation
	in three-dimensional hemodynamic simulations.
	\newblock {\em Comput. Methods Appl. Mech. Eng.} {\bf
		2020}, {\em 366},~113047.
	\newblock
	doi:\href{https://doi.org/10.1016/j.cma.2020.113047}{10.1016/j.cma.2020.113047}.
	
	\bibitem[Liu \em{et~al.}(2021)Liu, Pareschi, and Zhu]{liu2021bi}
	Liu, L.; Pareschi, L.; Zhu, X.
	\newblock A bi-fidelity stochastic collocation method for transport equations
	with diffusive scaling and multi-dimensional random inputs.
	\newblock {\em arXiv} {\bf 2021},  arXiv:2107.09250.
	
	\bibitem[Xiu(2010)]{XiuUQ}
	Xiu, D.
	\newblock {\em {Numerical Methods for Stochastic Computations}}; Princeton
	University Press, 2010.
	
	\bibitem[Zhu \em{et~al.}(2013)Zhu, Wu, Chern, and Sun]{Chern}
	Zhu, Y.; Wu, X.; Chern, I.L.; Sun, Z.
	\newblock {\em {Derivative Securities and Difference Methods}}, 2nd ed.;
	Springer,
	2013.
	
	\bibitem[Strikwerda(2004)]{StrikwerdaPDE}
	Strikwerda, J.C.
	\newblock {\em Finite Difference Schemes and Partial Differential Equations},
	2nd ed.; Society for Industrial and Applied Mathematics,
	2004.
	
\end{thebibliography}

\end{document}